\documentclass[aps,prd,twocolumn,groupedaddress,showpacs,showkeys,superscriptaddress]{revtex4-1}% all options

\usepackage{amsmath}
\usepackage{amssymb}
\usepackage{dsfont}
\usepackage{slashed}
\usepackage{color}
\usepackage[hidelinks]{hyperref}
\usepackage{graphics}
\usepackage{graphicx}
\usepackage{lipsum}
\usepackage{braket}
\usepackage{listings}

\newcommand{\Ho}{\textrm{H}}
\newcommand{\Ve}{\textrm{V}}
\newcommand{\Le}{\textrm{L}}
\newcommand{\Ri}{\textrm{R}}
\newcommand{\Di}{\textrm{D}}
\newcommand{\An}{\textrm{A}}
\newcommand{\T}{\textrm{T}}

\renewcommand{\vec}[1]{\boldsymbol{#1}}

\begin{document}

%\preprint{}

\title{Spin in Compton scattering with pronounced polarization dynamics}

\author{Sven Ahrens$^{*,}$}
\affiliation{Beijing Computational Science Research Center Zhongguancun Software Park II, No. 10 West Dongbeiwang Road, Haidian District, Beijing 100193, China}
\email{ahrens@csrc.ac.cn}
\author{Chang-Pu Sun}
\affiliation{Beijing Computational Science Research Center Zhongguancun Software Park II, No. 10 West Dongbeiwang Road, Haidian District, Beijing 100193, China}
\affiliation{Graduate School, China Academy of Engineering Physics, 100193 Beijing, China}
% \altaffilitation{Graduate School, China Academy of Engineering Physics}

\date{\today}

\begin{abstract}
We theoretically investigate a scattering configuration in Compton scattering, in which the orientation of the electron spin is reversed and simultaneously, the photon polarization changes from linear polarization into circular polarization. The intrinsic angular momentum of electron and photon are computed along the coincident propagation direction of the incoming and outgoing photon. We find that this intrinsic angular momentum is not conserved in the considered scattering process. We also discuss the generation of entanglement for the considered scattering setup and present an angle dependent investigation of the corresponding differential cross section, Stokes parameters and spin expectation.
\end{abstract}

\pacs{42.50.Ct,42.25.Ja,13.88.+e,03.67.Bg}
\keywords{Spin, Polarization, Electrons, Photons, Entanglement production}

% 42.50.Ct 	Quantum description of interaction of light and matter; related experiments
% 42.25.Ja 	Polarization
% 13.88.+e 	Polarization in interactions and scattering
% 03.67.Bg 	Entanglement production and manipulation

\maketitle

\section{Introduction}

The existence of electron spin effects in strong laser fields became interesting recently, as for example due to the experimental accessibility in strong-field ionization \cite{hartung_2016_spin_strong-field_ionization}, which is supported by theoretical investigations \cite{milosevic_2016_spin_polarized_electrons,Yakaboylu_2015_ATI_spin,klaiber_2014_spin_tunneling,klaiber_2014_spin_dynamics_HCI,barth_2014_hole_dynamics_and_spin_currents,barth_2013_spin-polarized_electrons,faisal_2004_spin_intense_field_ionization}. Spin-dependent weak-field ionization has also been investigated in the past \cite{Heckenkamp_1986_spin-resolved_photoelectron_spectroscopy,Dixit_1981_spin_polarization_of_electrons_in_ionization,Lambropoulos_1974_totally_polarized_electrons_through_multiphoton_ionization,Lambropoulos_1973_spin-orbit_coupling_and_photoelectron_polarization,Fano_1969_spin_orientation_of_photoelectrons} and even the existence of spin effects in strong fields without atoms was proposed \cite{ahrens_2016_electron_spin_filter,dellweg_mueller_extended_KDE_calculations,dellweg_mueller_2016_interferometric_spin-polarizer,dellweg_awwad_mueller_2016_spin-dynamics_bichromatic_laser_fields,erhard_bauke_2015_spin,McGregor_Batelaan_2015_two_color_spin,bauke_ahrens_2014_spin_precession_1,bauke_ahrens_2014_spin_precession_2,ahrens_bauke_2013_relativistic_KDE,ahrens_bauke_2012_spin-kde}. In the latter case an electron in vacuum interacts with the modes of the laser field and due to momentum conservation a diffraction pattern is formed. This property was first described theoretically by Kapitza and Dirac \cite{kapitza_dirac_1933_proposal} (see also for example \cite{batelaan_2007_RMP_KDE,fedorov_1974_adiabatic_switching,Schleich_2001_Quantum_Optics}) and has been demonstrated later experimentally with electrons \cite{Freimund_Batelaan_2002_KDE_detection_PRL,Freimund_Batelaan_2001_KDE_first,Bucksbaum_1988_electron_diffraction_regime} and also with atoms \cite{gould_1986_atoms_diffraction_regime,martin_1988_atoms_bragg_regime}.

Interestingly, the electron spin can alter in perpendicular direction to the laser propagation direction \cite{ahrens_bauke_2013_relativistic_KDE,ahrens_bauke_2012_spin-kde,bauke_ahrens_2014_spin_precession_1,bauke_ahrens_2014_spin_precession_2}, even though a famous proof by Wigner implies that photons only carry angular momentum along their propagation direction, ie. in longitudinal direction \cite{Wigner_1939_Unitary_Representations_of_Inhomogeneous_Lorentz_Group}. The question arises, how the electron may change its spin in the perpendicular direction, if the laser photons only carry longitudinal angular momentum.

Therefore we want to study the spin dynamics of the electron with polarized photons, because the above mentioned studies only consider a classical external field for the photonic sector. Within a simplification we ask for the spin dynamics for the interaction with a single photon field, for which the light-matter interaction simplifies to Compton scattering. Analytic expressions for spin-dependent Compton scattering were discussed by Franz  \cite{Franz_1938_streuung_von_strahlung} or Lipps and Tolhoek \cite{Lipps_Tolhoek_1954_polarization_phenomena_1,Lipps_Tolhoek_1954_polarization_phenomena_2} and in less generalized versions also by Fano \cite{Fano_1949_quantum-mechanical_treatment}, Klein and Nishina \cite{Klein_Nishina_1929_compton_streuung_1,Nishina_1929_compton_streuung_2} and Heitler \cite{heitler_1944_quantum_theory-of_radiation}. Further studies were discussing polarization in Compton scattering \cite{stedman_1982_circularly_polarized_compton_scattering,wightman_1948_polarized_compton_scattering}, its relations to the Stokes parameters \cite{bryne_1965_electron_polarization_stokes_vector,carrassi_1959_compton_scattering}, electron polarization for photon backscattering \cite{harutyunian_1964_electron_spin_polarization_from_backscattering} and electron polarization at high electron momenta \cite{kotkin_1998_high_electron_momenta_polarization}. The differential cross section for scattering with spin $1/2$ particles is discussed in \cite{anders_1983_compton_cross_section} and expressions for matrix elements in Compton scattering are given in \cite{hofri_1964_matrix_elements_compton_scattering}. Also a left-right asymmetry in Compton scattering is reported in \cite{bock_1971_left_right_assymetry_compton_scattering} and the author establishes relations between theory and experiment in \cite{bock_1971_compton_scattering}. Experimentally, polarization dynamics are investigated in \cite{raju_1968_experimental_polarized_compton_scattering} and spin polarization in Compton scattering is discussed in \cite{passchier_1998_compton_backscattering_spin_polarization} and \cite{mane_1997_spin_polarization_high_energy_accelerators}. A good overview about early works is also given in the review from Tolhoek \cite{tolhoek_1956_electron_polarization_review} and another overview on spin-dependent interaction of electrons with different targets is given in \cite{farago_1971_electron_spin_polarization_review}. Recently, also Compton scattering with twisted light has been investigated \cite{Stock_2015_Compton_scattering_of_twisted_light}. In strong fields, descriptions for the polarization dependent cross section have been studied in
\cite{karlovents_2011_strong_field_compton_scattering,Ivanov_2004_strong_field_compton_scattering,Bolshedvorsky_2000_spin_of_scattered_electrons,telnov_1995_principles_of_photon_colliders,Tsai_1993_equations_laser_particle_interactions,Grinchshin_Rekalo_1984,Galynskii_Sikach_1992_Nonlinear_effects_in_photon_emission,Baier_1975_quantum_processes_strong_electromagnetic_wave,Nikishov_Ritus_1963_quantum_processes}.

In a quantum mechanical description, coherent superpositions of different spin states may result in a new state with completely different spin alignment. Therefore, we are not only interested in the polarization direction but also in the phases of elements in the scattering matrix (S-matrix). Our investigation is based on the spin matrix structure of the S-matrix itself, which allows to deduce any polarization configuration of the outgoing particles in dependence of the incoming particles, for which even spin entanglement can be accounted for. We point out that the generation of entanglement is studied for example for the polarization of photons \cite{Aspect_1981_photon_entanglement}, internal electronic states of trapped ions \cite{Turchette_1998_trapped_ion_entanglement}, the spin of electron-hole pairs in solids \cite{Beenakker_2005_electron_hole_pair_spin_entanglement}, charge qubits in superconducting nanocircuits \cite{He_2003_charge_qbit_entanglement} or polarization and orbital angular momentum of photons \cite{Bhatti_spin_OAM_entanglement} (see also \cite{Nielsen_Chuang_2010_Quantum_computation})
and is therefore of interest for the scientific community.

We focus on a specific scattering process, in which the electron reverses its spin orientation while the photon polarization changes from linear into circular. We further compute the angular momentum of the spin polarizations and conclude that the spin angular momentum is not conserved for this process. This is interesting, because based on the assumption of spin conservation, arguments are given for spin-dependent selection rules in literature \cite{Batelaan_2003_MSGE}. In general, polarization transfer between charged particles and x-rays is relevant for polarization control of spin-polarized particle beams. Recent experimental studies of polarization transfer are \cite{blumenhagen_2016_x-ray_polarization_transfer,Martin_2012_polarization_transfer,weber_2011_Compton_polarimetry_monte_carlo,tashenov_2011_spin_polarization_transfer} and also theory investigations on this topic have been carried out in \cite{yerokhin_2010_theory_polarization_correlations,surzhykov_2005_ion_beam_spin_diagnostics}.

The paper is structured as follows. In section \ref{sec:compton_scattering} we introduce a Taylor expansion for a particular momentum configuration of the S-matrix in Compton scattering and emphasize a specific scattering process which is subject of investigation in this paper. We discuss the generation of entanglement in section \ref{sec:entanglement_generation} and analyze the spin of the electron and photon polarization in section \ref{sec:electron_and_photon_spin} for the specific scattering event. Section \ref{sec:angular_analysis} contains an angle dependent investigation of the differential cross section, Stokes parameters and spin expectation of the interaction. Finally we discuss how such a scattering can be explicitly observed in experiment in section \ref{sec:experimental_setup}.

\vfill

\section{Spin-dependent Compton scattering\label{sec:compton_scattering}}

Our studies are based on the S-matrix in Compton scattering. For this we refer to chapter 3.7 in reference \cite{greiner_reinhardt_1992_quantum_electrodynamics}, in which the S-matrix reads
\begin{widetext}
\begin{subequations}%
\begin{equation}%
 S_{fi} = -i \frac{q^2}{V^2} \sqrt{\frac{m^2}{\mathcal{E}_i \mathcal{E}_f}} \sqrt{\frac{(4 \pi)^2}{2 \omega 2 \omega'}} (2 \pi)^4 \delta^4(p_f + k' - p_i - k)\\
 \cdot \epsilon^\mu(\vec k') \epsilon^\nu(\vec k) M_{\mu\nu}(s_f,s_i)\frac{1}{m}\,, \label{eq:S-matrix}
\end{equation}%
with Compton tensor%
\begin{equation}%
  M_{\mu\nu}(s_f,s_i) = m \ \bar u(\vec p_f,s_f) \left[ \gamma_\mu \frac{\slashed p_i + \slashed k + m}{2 p_i \cdot k} \gamma_\nu + \gamma_\nu \frac{\slashed p_i - \slashed k' + m}{-2 p_i \cdot k'} \gamma_\mu \right] u(\vec p_i,s_i)\,.\label{eq:compton_tensor}
\end{equation}\label{eq:full_S-matrix}%
\end{subequations}%
\end{widetext}
Here, $q$ is the electron charge, $m$ is the electron restmass and $V$ is a normalization volume. The speed of light $c$ and the reduced Planck constant $\hbar$ are set to unity in the Gaussian unit system and we use the Feynman slash notation with Dirac gamma matrices. The polarizations $\epsilon^\nu(\vec k)$ and $\epsilon^\mu(\vec k')$ are the part of the photon's vector potential
\begin{equation}
 A^\nu(x,k)=\sqrt{\frac{4 \pi}{2 \omega V}} \left( \epsilon^\nu (\vec k) e^{-i k \cdot x} + \epsilon^{\nu*} (\vec k) e^{i k \cdot x} \right)\label{eq:incoming-photon-field}
\end{equation}
and the four component Dirac spinor $u(p,s)$ is given in Ref. \cite{greiner_2000_relativistic_quantum_mechanics}. $\mathcal{E}_i$ and $\mathcal{E}_f$ are the relativistic energy momentum relations of the incoming and outgoing electrons with momenta $\vec p_i$ and $\vec p_f$ and $\mathcal{\omega}$ and $\mathcal{\omega}'$ are the photon frequencies of the incoming and outgoing photons with momenta $\vec k$ and $\vec k'$, respectively.

Next we want to investigate the electron-photon interaction for the case that the electron undergoes a spinflip and know that this certainly happens, when the electron transverses a linear polarized, standing wave of light with electron momentum $m$ along the standing wave's polarization direction \cite{ahrens_bauke_2013_relativistic_KDE}. Assuming that the electron has picked up and emitted a photon out of the standing wave of light leads us to the initial and final photon momenta
\begin{subequations}%
\begin{align}%
 \vec k  & = \phantom{-}k_p \vec e_x \textrm{ and }\label{eq:initial_photon_momentum}\\
 \vec k' & = -k_p \vec e_x
\end{align}\label{eq:photon_momenta}%
\end{subequations}%
respectively, for the single photon interaction analogon considered here. $k_p$ is the photon momentum. The electron is required to enter and leave the scattering region at the Bragg angle, with initial and final momentum
\begin{subequations}%
\begin{align}%
 \vec p_i&=-k_p \vec e_x + p_2 \vec e_y + p_3 \vec e_z \label{eq:initial_electron_momentum}\\
 \vec p_f&=\phantom{-}k_p \vec e_x + p_2 \vec e_y + p_3 \vec e_z\,,
\end{align}%
\end{subequations}%
respectively, which to fulfills energy and momentum conservation. In accordance with the geometry in Ref. \cite{ahrens_bauke_2013_relativistic_KDE} we want to study the scattering around the parameters $p_2 = 0$ and $p_3 \approx m$.

A Taylor expansion of the Compton tensor \eqref{eq:compton_tensor} up to first order with respect to the small, dimensionless parameters $\beta=p_2/m$ and $\gamma=1 -p_3/m$ and up to second order in the small parameter $\alpha=k_p/m$ yields
% \begin{subequations}
% \begin{align}
%  M_{22} &= \frac{1}{m} \,\mathds{1} +i \frac{1 -\sqrt{2}}{\sqrt{2}} \frac{k_p}{m^2} \sigma_y + \frac{-1 + \sqrt{2}}{2} \frac{k_p^2}{m^3} \mathds{1} \label{eq:M_22} \\
%  M_{23} &= -i\frac{1}{2} \frac{k_p}{m^2}\sigma_x -i\frac{1}{2} \frac{k_p}{m^2}\sigma_z - \frac{p_2}{m^2} \mathds{1}  \label{eq:M_23}\\
%  M_{32} &= \phantom{-} i\frac{1}{2} \frac{k_p}{m^2}\sigma_x -i\frac{1}{2} \frac{k_p}{m^2}\sigma_z - \frac{p_2}{m^2} \mathds{1} \label{eq:M_32}\\
%  M_{33} &=i \frac{1}{\sqrt{2}} \frac{k_p}{m^2} \sigma_y + \frac{m - p_3}{m^2}\mathds{1} + \frac{-1 + \sqrt{2}}{2} \frac{k_p^2}{m^3} \mathds{1}\,. \label{eq:M_33}
% \end{align}\label{eq:taylor_expansion_compton_tensor}
% \end{subequations}
\begin{subequations}%
\begin{align}%
 M_{22} &= \mathds{1} +i \frac{1 -\sqrt{2}}{\sqrt{2}} \alpha\, \sigma_y + \frac{-1 + \sqrt{2}}{2} \alpha^2 \mathds{1} \label{eq:M_22} \\
 M_{23} &= \phantom{\mathds{1}-(q_2}-i\frac{1}{2} \alpha\,(\sigma_x + \sigma_z) - \beta\, \mathds{1} \label{eq:M_23}\\
 M_{32} &= \phantom{,+\mathds{1}-(q_2} i\frac{1}{2} \alpha\,(\sigma_x - \sigma_z ) - \beta\, \mathds{1} \label{eq:M_32}\\
 M_{33} &= \phantom{,\mathds{1}-(q_2} i \frac{1}{\sqrt{2}} \alpha\, \sigma_y +  \gamma\, \mathds{1} + \frac{-1 + \sqrt{2}}{2} \alpha^2 \mathds{1} \,. \label{eq:M_33}
\end{align}\label{eq:taylor_expansion_compton_tensor}%
\end{subequations}%
Here, $\mathds{1}$ is the $2\times 2$ identity matrix, $\sigma_i$ are the Pauli matrices and the combined object
\begin{equation}
 M=
 \begin{pmatrix}
  M_{22} & M_{23} \\
  M_{32} & M_{33}
 \end{pmatrix}
\end{equation}
is a $4\times 4$ matrix, which consists of the $2\times2$ matrices
\begin{equation}
 M_{\mu\nu} =
\begin{pmatrix}
 M_{\mu\nu}(s^\uparrow,s^\uparrow) &  M_{\mu\nu}(s^\uparrow,s^\downarrow) \\
 M_{\mu\nu}(s^\downarrow,s^\uparrow) &  M_{\mu\nu}(s^\downarrow,s^\downarrow)
\end{pmatrix}\,.
\end{equation}
We denote this in the following way by using the Dirac bra- and ket notation. The initial and final electron spin $s_i$ and $s_f$ can assume the  quantization directions spin-up $s^\uparrow=(1,0)^\T$ or spin-down $s^\downarrow=(0,1)^\T$. The initial and final photon polarization $\epsilon(\vec k)^\mu$ and $\epsilon(\vec k')^\nu$ assumes the horizontal and vertical polarization states 
\begin{equation}
 \epsilon^\Ho = \vec e_y\,, \qquad \epsilon^\Ve = \vec e_z\,,
\end{equation}
which in turn are basis states for left and right circular polarization
\begin{equation}
 \epsilon^\Le = \frac{1}{\sqrt{2}}\left(\epsilon^\Ho + i \epsilon^\Ve\right) \,, \qquad \epsilon^\Ri = \frac{1}{\sqrt{2}}\left(\epsilon^\Ho - i \epsilon^\Ve\right) \label{eq:circular_polarization}
\end{equation}
and for diagonal and anti-diagonal polarization
\begin{equation}
 \epsilon^\Di = \frac{1}{\sqrt{2}}\left(\epsilon^\Ho + \epsilon^\Ve\right) \,, \qquad \epsilon^\An = \frac{1}{\sqrt{2}}\left(\epsilon^\Ho - \epsilon^\Ve\right) \,.\label{eq:diagonal_polarization}
\end{equation}
Then Eq. \eqref{eq:taylor_expansion_compton_tensor} consists of 16 matrix elements $\bra{\epsilon_f,s_f}M\ket{\epsilon_i,s_i}$. The photon polarization $\epsilon_i,\epsilon_f \in \{\Ho,\Ve\}$ is indexed explicitly by the numbers $\{2,3\}$ and the spin polarization $s_f,s_i \in \{\uparrow,\downarrow\}$ is implied by the $2\times 2$ matrix structure of each of the four subequations in Eq. \eqref{eq:taylor_expansion_compton_tensor}. The tensor product of the photon and electron polarization $\ket{\epsilon_i,s_i} = \ket{\epsilon_i}\otimes\ket{s_i}$ forms a basis with the four basis states $\{\Ho\uparrow,\Ho\downarrow,\Ve\uparrow,\Ve\downarrow\}$ with respect to which the initial combination of photon and electron spin configuration can be expanded to
\begin{equation}
 \ket{\psi_i}= \varphi_1\ket{\Ho,\uparrow}+\varphi_2\ket{\Ho,\downarrow}+\varphi_3\ket{\Ve,\uparrow}+\varphi_4\ket{\Ve,\downarrow}\,.
\end{equation}
In this sense, Eq. \eqref{eq:taylor_expansion_compton_tensor} shows the matrix elements, which relate the initial product state of photon and electron spin $\ket{\psi_i}$ to the final product state
\begin{equation}
 \ket{\psi_f} = \sum_{\epsilon_f \in \{\Ho,\Ve\}\atop s_f \in \{\uparrow,\downarrow\}} \ket{\epsilon_f,s_f} \braket{\epsilon_f,s_f|M|\psi_i}\label{eq:in-out-relation}
\end{equation}
of the outgoing photon and electron in the case of scattering.

\begin{figure}[t]%
  \includegraphics[width=0.48\textwidth]{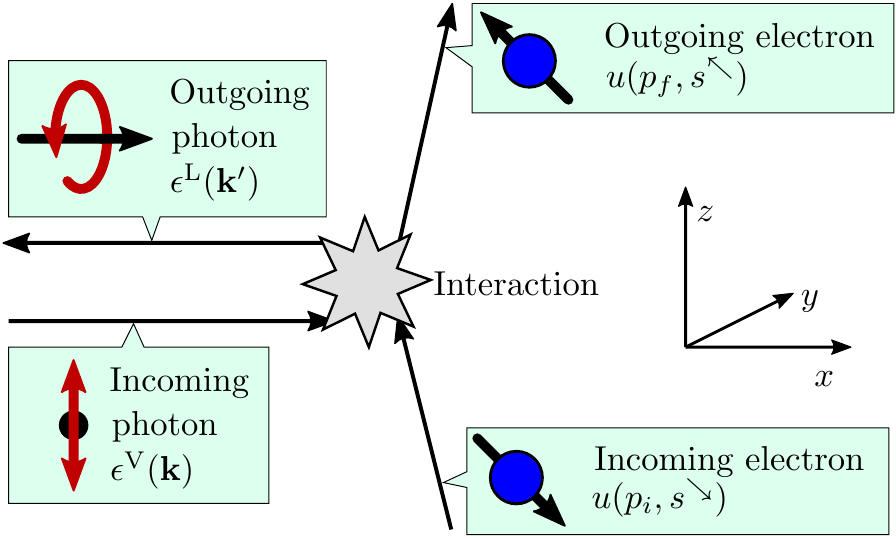}%
  \caption{\label{fig:electron-photon_interaction}%
(Color online) Illustration of the considered process \eqref{eq:Vse-Lnw_scattering}, in which electron and photon are changing their spin-state simultaneously. The vertically polarized photon with momentum $k_p$ in $x$ direction and the electron with momentum $-k_p \vec e_x + p_3 \vec e_z$ with $p_3\approx m$ and spin orientation $(1,0,-1)^\T/\sqrt{2}$
 are entering the interaction region. We consider a scattering process in which both particles reverse their momentum component in $x$ direction. According to Eq. \eqref{eq:Vse-Lnw_scattering} the outgoing photon is left-circularly polarized and the spin of the outgoing electron spin is pointing in $(-1,0,1)^\T/\sqrt{2}$ direction after the interaction.}%
\end{figure}

The leading terms in Eq. \eqref{eq:taylor_expansion_compton_tensor} scale with $\alpha$, where the other terms proportional to $\beta$, $\gamma$ and $\alpha^2$ can be assumed to be negligible for the momenta $p_i$, $p_f$, $k$ and $k'$, which are of interest here. A full spinflip of the electron will occur, if the incoming photon is vertically polarized, because the spin-preserving term $\mathds{1}$ in $M_{22}$ only arises for horizontal polarization. For a vertical initial photon polarization a spin-dependent term proportional to $\sigma_x + \sigma_z$ arises from $M_{23}$ and a spin-dependent term proportional to $\sigma_y$ arises from $M_{33}$. The corresponding directions $(1,0,1)^\T/\sqrt{2}$ and $(0,1,0)^\T/\sqrt{2}$ have the orthogonal direction $(1,0,-1)^\T/\sqrt{2}$. The eigenvectors of the spin-dependent term $\sigma_x - \sigma_z$, which corresponds to the $(1,0,-1)^\T/\sqrt{2}$ direction, will be flipped when acting on the spin terms in $M_{22}$ and $M_{33}$. Therefore, by using the Bloch state
\begin{equation}
 s(n) =
\begin{pmatrix}
 \cos \frac{\theta_n}{2} \\ \sin \frac{\theta_n}{2} e^{i \varphi}
\end{pmatrix}
\end{equation}
with $\theta_n=2\pi + (1+ 2n)\pi/4$ and $\varphi = 0$ we define
\begin{equation}
 s^\nearrow=s(0)\,,\ s^\searrow=s(1)\,,\ s^\swarrow=s(2)\,,\ s^\nwarrow=s(3)\,,\label{eq:tilted_spinor_definition}
\end{equation}
together with the eigenstates of $\sigma_y$
\begin{equation}
 s^\otimes=\frac{1}{\sqrt{2}}
 \begin{pmatrix}
 1 \\ i
 \end{pmatrix}
 \,,\qquad
 s^\odot=\frac{1}{\sqrt{2}}
 \begin{pmatrix}
 1 \\ - i
 \end{pmatrix}\,.
\end{equation}
The set of spinors $\{s^\swarrow,s^\nearrow,s^\otimes,s^\odot,s^\searrow,s^\nwarrow\}$ are 45 degrees tilted around the $y$ axis as compared to the commonly normalized eigensolutions of the $\sigma_x$, $\sigma_y$ and $\sigma_z$ Pauli matrices. Accordingly, one obtains the spin expectation values
\begin{subequations}%
\begin{align}%
 s^{\swarrow\dagger} \vec \sigma s^\swarrow &= (-1,0,-1)^\T/\sqrt{2} \\
 s^{\nearrow\dagger} \vec \sigma s^\nearrow &= (1,0,1)^\T/\sqrt{2} \\
 s^{\otimes\dagger} \vec \sigma s^\otimes &= (0,1,0)^\T \\
 s^{\odot\dagger} \vec \sigma s^\odot &= (0,-1,0)^\T \\
 s^{\searrow\dagger} \vec \sigma s^\searrow &= (1,0,-1)^\T/\sqrt{2} \\
 s^{\nwarrow\dagger} \vec \sigma s^\nwarrow &= (-1,0,1)^\T/\sqrt{2}
\end{align}\label{eq:spin-expectation_values}%
\end{subequations}%
and the spinors $s^\swarrow$, $s^\otimes$, $s^\searrow$ are eigenvectors of $(\sigma_x + \sigma_z)/\sqrt{2}$, $\sigma_y$ and $(\sigma_x - \sigma_z)/\sqrt{2}$ with eigenvalue 1 and $s^\nearrow$, $s^\odot$, $s^\nwarrow$ are eigenvectors with eigenvalue -1, respectively. In numeric implementations we use an equivalent expression for $s^\searrow$ and $s^\nwarrow$ given in appendix \ref{sec:spinor_expressions}. For ease of interpretation we want to transform the electron spin degree of freedom into the quantization axis $s^\searrow$ and $s^\nwarrow$ by the electron spin rotation $M_{ab}'=U^\dagger M_{ab} U,\, a,b\in\{2,3\}$ with the matrix
\begin{equation}
 U=  \left( s^\searrow,s^\nwarrow \right) =
 \begin{pmatrix}
  \cos \frac{\theta_1}{2} & \cos \frac{\theta_3}{2} \\
  \sin \frac{\theta_1}{2} e^{i \varphi} & \sin \frac{\theta_3}{2} e^{i \varphi} \\
 \end{pmatrix} \,.\label{eq:unitary_spin_transformation}
\end{equation}
With respect to this quantization axis and by neglecting the terms proportional to $\beta$, $\gamma$ and $\alpha^2$ the transformed matrix elements of $M'$ read
% \begin{align}
%  M_{22}' &= \frac{1}{m} \,\mathds{1} & 
%  M_{23}' &= i \frac{1}{\sqrt{2}} \frac{k_p}{m^2} \sigma_x \nonumber \\
%  M_{32}' &= i \frac{1}{\sqrt{2}} \frac{k_p}{m^2} \sigma_z &
%  M_{33}' &=i \frac{1}{\sqrt{2}} \frac{k_p}{m^2} \sigma_y \,.
% \end{align}
\begin{align}
 M_{22}' &= \mathds{1} & 
 M_{23}' &= i \frac{1}{\sqrt{2}} \alpha\, \sigma_x \nonumber \\
 M_{32}' &= i \frac{1}{\sqrt{2}} \alpha\, \sigma_z &
 M_{33}' &=i \frac{1}{\sqrt{2}} \alpha\, \sigma_y \,.
\end{align}
The matrices $M'$ imply that a vertically polarized photon and a $s^\searrow$ polarized electron, denoted by the product state $\ket{\Ve,\searrow}$, will be scattered into a left-circularly polarized photon and a $s^\nwarrow$ polarized electron, denoted by $\ket{\Le,\nwarrow}$. This process can be written as
\begin{subequations}%
\begin{equation}%
 M \ket{\Ve,\searrow} = i \frac{k_p}{m} \ket{\Le,\nwarrow}\,.\label{eq:Vse-Lnw_scattering}
\end{equation}%
The scattering process of the electron and photon in Eq. \eqref{eq:Vse-Lnw_scattering}  and their spin properties are illustrated in Fig. \ref{fig:electron-photon_interaction}.%

Similarly, a vertically polarized photon and a $s^\nwarrow$ polarized electron, denoted by $\ket{\Ve,\nwarrow}$ will be scattered into a right-circularly polarized photon and a $s^\searrow$ polarized electron, denoted by $\ket{\Ri,\searrow}$%
\begin{equation}%
 M \ket{\Ve,\nwarrow} = i \frac{k_p}{m} \ket{\Ri,\searrow}\,.\label{eq:Vse-Rnw_scattering}
\end{equation}\label{eq:V_scattering}%
\end{subequations}%

\section{Generation of entanglement\label{sec:entanglement_generation}}

A coherent superposition of the incoming states in the processes \eqref{eq:V_scattering} can be used for establishing entanglement between the electron spin and the photon polarization. The spin state $s^\swarrow$ can be represented as a superposition of the spinors $s^\searrow$ and $s^\nwarrow$ by $s^\swarrow = (s^\searrow + s^\nwarrow)/\sqrt{2}$. On the other hand, the electron spin of $s^\swarrow$ points perpendicularly to the directions of $s^\searrow$ and $s^\nwarrow$, as can be seen from their spin expectation values \eqref{eq:spin-expectation_values}. Due to the superposition property of the spinor $s^\swarrow$ one can write
\begin{subequations}%
\begin{align}%
 M\ket{\Ve,\swarrow} &= M \frac{1}{\sqrt{2}}\left( \ket{\Ve,\searrow} + \ket{\Ve,\nwarrow} \right)\\
 &= i \frac{k_p}{m} \frac{1}{\sqrt{2}} \left( \ket{\Le,\nwarrow} + \ket{\Ri,\searrow} \right)\,.\label{eq:ssw_scattered_result}
\end{align}\label{eq:ssw_scattereding}%
\end{subequations}%
The second line of Eq. \eqref{eq:ssw_scattereding} shows that the scattering of a vertically polarized photon with a $s^\swarrow$ polarized electron results in an entangled superposition of a left circularly polarized photon with a $s^\nwarrow$ polarized electron and a right circularly polarized photon with a $s^\searrow$ polarized electron, which is a maximally entangled (Bell) state \cite{Nielsen_Chuang_2010_Quantum_computation}.

\section{Assigning spin to the polarization\label{sec:electron_and_photon_spin}}

\subsection{Intrinsic angular momentum density of electron\label{sec:electron_spin}}

For the process \eqref{eq:Vse-Lnw_scattering} we investigate the intrinsic angular momentum density (spin density) of the photon and electron before and after scattering. For the electron spin density different relativistic spin-operators have been considered in the literature, among which only the proposals from Foldy and Wouthuysen $\vec {\hat S}_\textrm{FW}$ as well as the proposal from Pryce $\vec {\hat S}_\textrm{Pr}$ fulfill the angular momentum algebra, are self-adjoint and have spectrum $\pm 1/2$ \cite{bauke_ahrens_keitel_grobe_2014_relativistic_spin_operators,bauke_ahrens_keitel_grobe_2014_relativistic_spin_operators_2}. Both operators are identical for the positive solutions of the free Dirac equation
%and the spin
and one can show that
\begin{equation}
%S_e(p,s)=
\frac{m}{V \mathcal{E}} \int_V d^3 x \,e^{ip\cdot x} u(p,s)^\dagger \hat{\vec S}_\textrm{FW} u(p,s) e^{-ip\cdot x}
\end{equation}
has the value $(\sin \theta \cos \varphi,\cos \theta \sin \varphi,\cos \theta)^\T/2$, if $s$ was the Bloch state $s=(\cos \theta/2, e^{i \varphi} \sin \theta/2 )^\T$. Therefore we conclude that the incoming electron in Eq. \eqref{eq:Vse-Lnw_scattering} has intrinsic angular momentum $(1,0,-1)^\T/\sqrt{8}$ and the outgoing electron has intrinsic angular momentum $(-1,0,1)^\T/\sqrt{8}$. The factor $\sqrt{8}$ in the denominator results from the normalization of the vector of the electron's spin expectation value and the property that the electron is a spin 1/2 particle. Thus, the total spin angular momentum $(-1,0,1)^\T/\sqrt{2}$ is transferred to the electron.

\subsection{Intrinsic angular momentum density of photon\label{sec:photon_spin}}

For the photon we consider the spin density $\vec E \times \vec A$, with $\vec E=-\dot{\vec A^\perp}$ and $\vec A^\perp$ being the transverse part of the vector potential $\vec A$ for plane waves \cite{mandel_wolf_1995_quantum_optics_book,Cameron_2014_second_potential_in_electrodynamics},\footnote{The spin-density $\vec E \times \vec A$ may depend on a duality transformation of the fields \cite{Barnett_2010_rotation_of_electromagnetic_fields,Bliokh_2013_dual_electromagnetism}. Such a duality transformation introduces new fields in the Lagrangian \cite{Cameron_2012_electric-magnetic_symmetry} and a generalization of the interacting theory of electrons and photons would be required for the S-matrix in \eqref{eq:full_S-matrix}. This is beyond the scope of this publication and therefore we base our conclusions on the spin-density $\vec E \times \vec A$ only.}\,. The given photon spin density is a gauge-dependent quantity. However, only the 0-component and the longitudinal component $\vec A^\parallel$ of the four vector potential of a plane wave depend on gauge \cite{Cameron_2014_second_potential_in_electrodynamics}. Since the electric field $\vec E$ of the free Maxwell equations only has a transverse component, the gauge-dependent part $\vec E \times \vec A^\parallel$ of the spin-density $\vec E \times \vec A$ is transverse as well. The longitudinal component of the spin-density $\vec E \times \vec A$ is therefore a gauge-invariant quantity. For this reason we only consider the longitudinal component of the photon spin density and mention that a photon (or in general massless particles) may only carry intrinsic angular momentum along its longitudinal direction, as Wigner has concluded in 1939 by studying the transformation properties of particles \cite{Wigner_1939_Unitary_Representations_of_Inhomogeneous_Lorentz_Group},\footnote{Note, that photons inside media may also have a transverse component of angular momentum \cite{Bliokh_Nori_2015_Transverse_and_longitudinal_angular_momenta_of_light}.}. The longitudinal $x$ component of the intrinsic photon angular momentum
\begin{equation}
 \frac{1}{4 \pi} \int_V d^3 x \left[\vec E(\vec x,t) \times \vec A(\vec x,t) \right]\cdot \vec e_x \label{eq:photon_spin}
\end{equation}
with the vector potential \eqref{eq:incoming-photon-field} 
yields the photon spin 0 for a vertically polarized incoming photon, the photon spin $1$ for a left-circularly polarized outgoing photon and the photon spin $-1$ for a right-circularly polarized outgoing photon.

The $x$ component of the spin of the electron and photon for the process \eqref{eq:Vse-Lnw_scattering} is summarized in table \ref{tab:spin_summary}, which contains the information that the net amount of $1 - 1/\sqrt{2}$ of spin (in units of $\hbar$) is not transported via the polarization  degree of freedom of the particles.

\begin{table}[]
\caption{Spin $x$ component of particles in the scattering process Eq. \eqref{eq:Vse-Lnw_scattering}. The table lists the $x$ component of the spin of the electron according to section \ref{sec:electron_spin} and the spin of the photon according to section \ref{sec:photon_spin}. The total spin is the sum of the spin of each particle species and the net spin transfer is the difference between outgoing and incoming spins. The lower right entry implies that a spin discrepancy of $1 - 1/\sqrt{2}$ (in units of $\hbar$) remains when the scattering occurs.\label{tab:spin_summary}}
\begin{center}
\begin{tabular}{ l | c | c || c }
  \ Spin of & \ Photon\,\, & \ Electron\,\, & \ Total\,\, \\ \hline
  \ Incoming particles\,\, & 0 & $1/\sqrt{8}$ & $1/\sqrt{8}$ \\ \hline
  \ Outgoing particles\,\, & 1 & $-1/\sqrt{8}$ & $1 - 1/\sqrt{8}$ \\ \hline \hline
  \ Net spin transfer\,\, & 1 & $-1/\sqrt{2}$ & $1 - 1/\sqrt{2}$
\end{tabular}\\
\end{center}
\end{table}

\section{Angular dependent scattering analysis\label{sec:angular_analysis}}

In this section we investigate scattering for general outgoing particle momenta. We denote the outgoing photon momentum by
\begin{equation}
 \vec k'(\vartheta,\varphi) = \omega'(\vartheta,\varphi)
\begin{pmatrix}
 \cos \vartheta \\ \sin \vartheta \cos \varphi \\ \sin \vartheta \sin \varphi
\end{pmatrix}\,, \label{eq:kp}
\end{equation}
where the parameter $\vartheta$ is the scattering angle and
\begin{equation}
 \omega'(\vartheta,\varphi) = k_p \left( 1 - \frac{p_3 \sin \vartheta \sin \varphi}{\mathcal{E}_i + k_p} \right)^{-1}\,,
\end{equation}
is the photon energy, which is implied by four momentum conservation
\begin{equation}
 p_i + k = p_f + k'\,,\label{eq:four_momentum_conservation}
\end{equation}
as computed in appendix \ref{sec:differential_cross_section}.
Correspondingly, the outgoing electron momentum is
\begin{equation}
 \vec p_f =
 \begin{pmatrix}
 -\omega'(\vartheta,\varphi) \cos \vartheta \\ - \omega'(\vartheta,\varphi) \sin \vartheta \cos \varphi \\ p_3 - \omega'(\vartheta,\varphi) \sin \vartheta \sin \varphi
\end{pmatrix}\,. \label{eq:final_electron_momentum}
\end{equation}

\subsection{Differential cross section}

The differential cross section for a process with incoming momenta $p_i$, $k$ and outgoing momenta $p_f$, $k'$ is
\begin{multline}
 \frac{d \sigma}{d \Omega}(\epsilon_f,s_f) = \frac{\alpha^2}{m^2} \frac{\omega^{\prime2}}{k_p^2} \frac{m}{\sqrt{\mathcal{E}_i^2 - 2 \mathcal{E}_i k_p + k_p^2 + p_3^2}} \\ \cdot \frac{m}{\mathcal{E}_i + k_p} \left|\braket{\epsilon_f,s_f|M|\psi_i}\right|^2\,, \label{eq:differential_cross_section}
\end{multline}
where the prefactor in front of the absolute square of the Compton tensor is derived in appendix \ref{sec:differential_cross_section}. The number $\alpha$ is the fine structure constant.

\begin{figure}[!h]%
  \includegraphics[width=0.48\textwidth]{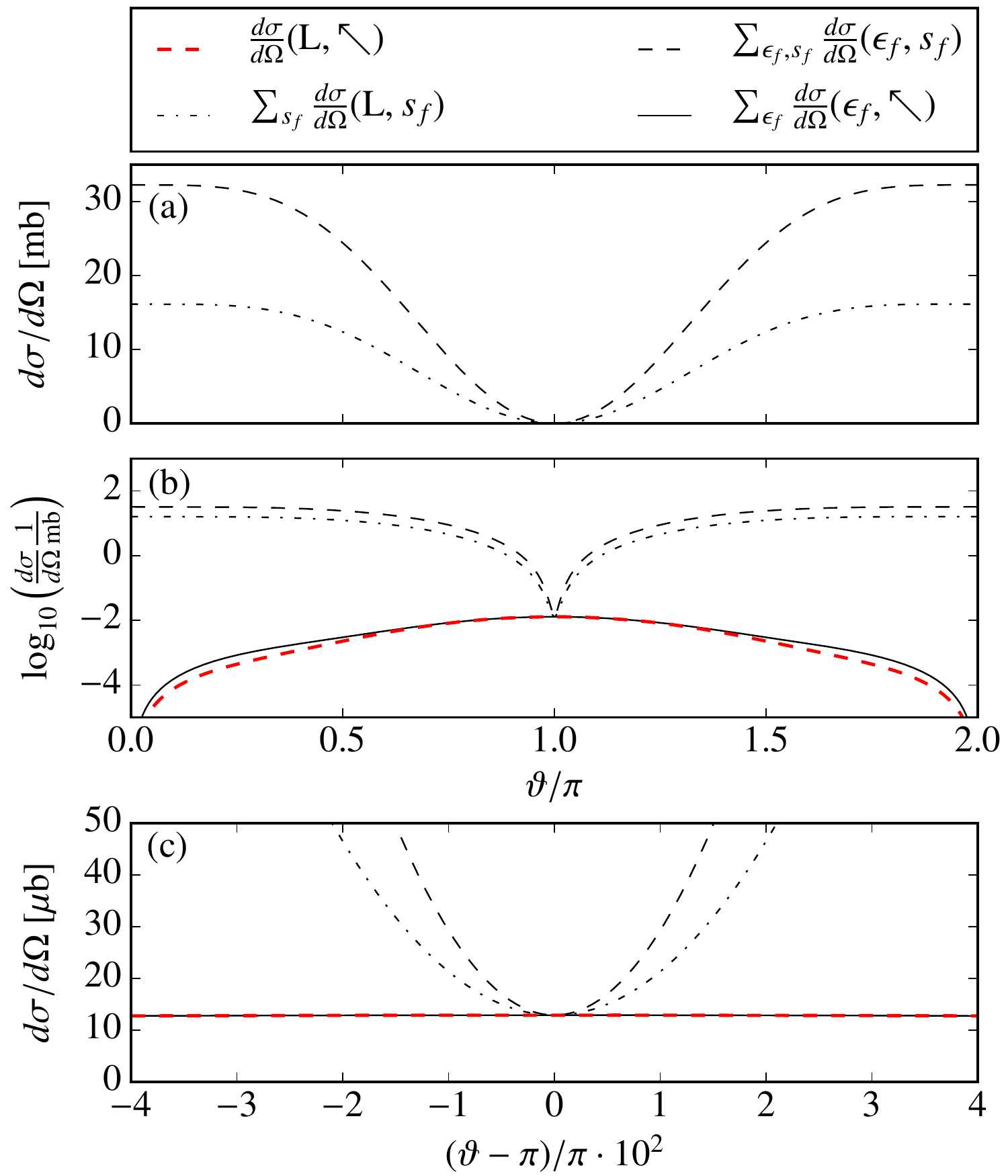}%
  \caption{(Color online) Plot of the differential cross section \eqref{eq:differential_cross_section} in the $x$-$y$ plane, which is shown as function of the scattering angle $\vartheta$ of the final photon momentum \eqref{eq:kp} with $\varphi=0$. The corresponding basis of the photon polarization vectors is given in \eqref{eq:photon_polarization_x_y}, with left and right circular polarization as introduced in Eq. \eqref{eq:circular_polarization}. The polarization sums run over the photon polarizations $\epsilon_f \in \{ \Le, \Ri \}$ and the spin sums run over the spin states $s_f \in \{ \searrow,\nwarrow \}$. The cross section for observing the final polarization state $\ket{\Le,\nwarrow}$ (red dashed line) and the polarization summed cross section (black solid line) are suppressed as compared to the spin summed cross section (black dashed dotted line) and the polarization and spin summed cross section (black dashed line). This is the reason, why the plot in panel (a) is repeated as logarithmic plot in panel (b). All plotted cross sections approach the same value in a dip at scattering angle $\vartheta=\pi$, which can be seen in a magnified plot in panel (c)\label{fig:differential_cross_section_x_y}
}%
\end{figure}

First we want to investigate the differential cross section \eqref{eq:differential_cross_section} for a photon which is scattered into the $x$-$y$ plane. To achieve this, we set $\varphi=0$ in Eq. \eqref{eq:kp}. With this constraint we plot the differential cross section over the interval $\vartheta \in [0,2\pi]$ in Fig. \ref{fig:differential_cross_section_x_y},
where the two photon polarizations
\begin{subequations}%
\begin{align}%
 \epsilon^{\prime\Ho}(\vartheta) &= \sin \vartheta \,\vec e_x - \cos \vartheta \,\vec e_y\\
 \epsilon^{\prime\Ve}(\vartheta) &= \vec e_z\,,
\end{align}\label{eq:photon_polarization_x_y}%
\end{subequations}%
are co-aligned to the final photon momentum \eqref{eq:final_electron_momentum} such that
\begin{equation}
 \vec k' \cdot \vec \epsilon^{\prime\Ve} = 0\,, \quad
 \vec k' \cdot \vec \epsilon^{\prime\Ho} = 0 \quad\textrm{and}\quad
 \vec \epsilon^{\prime\Ve} \cdot \vec \epsilon^{\prime\Ho} = 0\label{eq:polarization_vector_orthogonality}
\end{equation}
is fulfilled. Circular polarizations $\epsilon^{\Le}$ and $\epsilon^{\Ri}$ and diagonal polarizations $\epsilon^\Di$ and $\epsilon^\An$ are related to $\epsilon^{\Ho}$ and $\epsilon^{\Ve}$ by the relations in Eq. \eqref{eq:circular_polarization} and \eqref{eq:diagonal_polarization}.
We point out that for the scattering angle $\vartheta=\pi$ the definitions for $k'$, $p_f$, $\epsilon^{\prime\Ve}$ and $\epsilon^{\prime\Ho}$ are coincident with the definitions made in section \ref{sec:compton_scattering}. In Fig. \ref{fig:differential_cross_section_x_y}, we see that a final polarization and spin configuration different than $\ket{\Le,\nwarrow}$ is dominating interactions over a wide range of the scattering angle $\vartheta$. However, the polarization and spin summed cross section coincides with the $\ket{\Le,\nwarrow}$ projected cross section at scattering angle $\vartheta=\pi$, when the electron is scattered backwards by 180 degrees. This implies that the polarization state of the outgoing particles must be in the state $\ket{\Le,\nwarrow}$ at scattering angle $\vartheta=\pi$ and confirms the conclusions from section \ref{sec:compton_scattering}.

\begin{figure}[!h]%
  \includegraphics[width=0.48\textwidth]{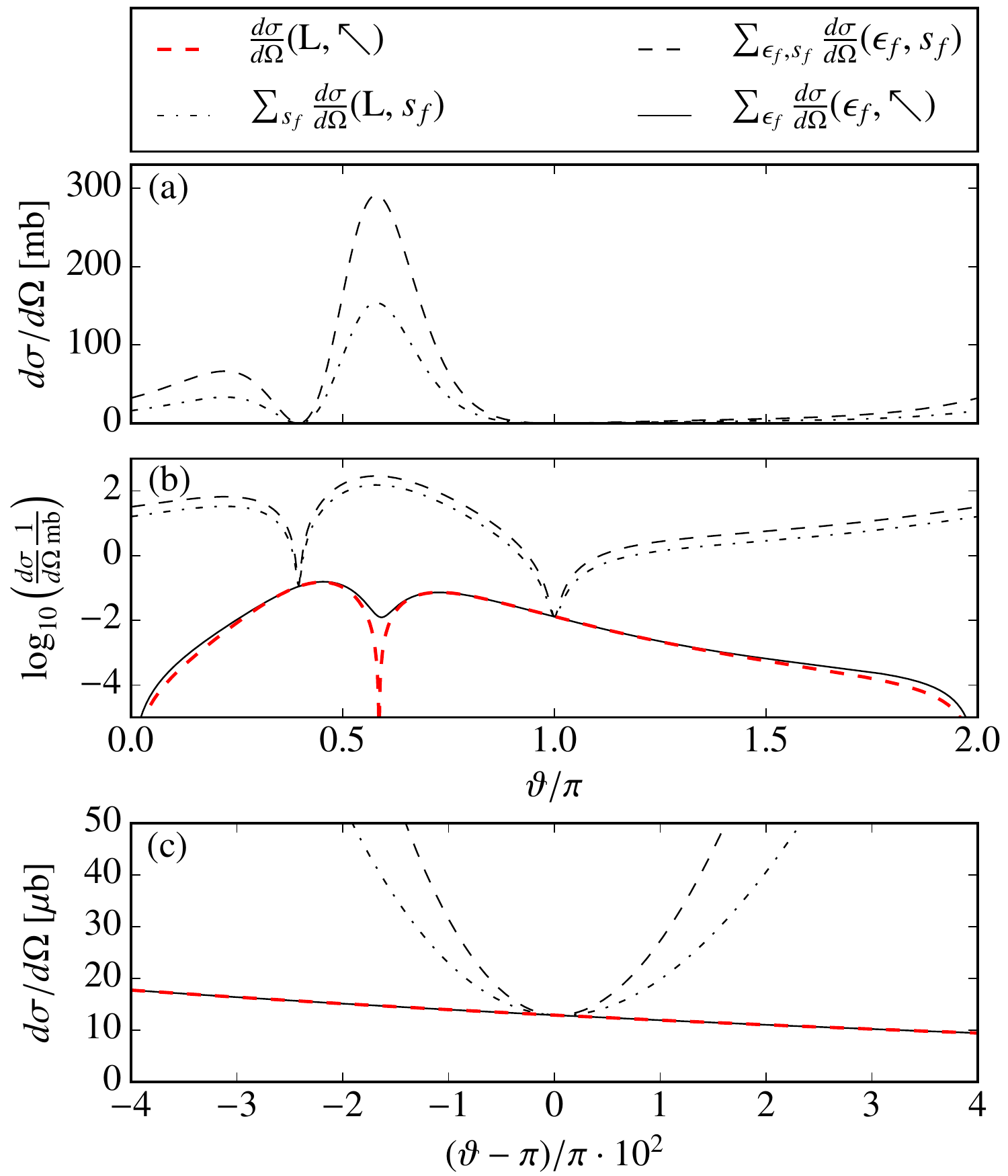}%
  \caption{(Color online) Plot of the differential cross section \eqref{eq:differential_cross_section} in the $x$-$z$ plane, as function of the scattering angle $\vartheta$ with $\varphi=\pi/2$ and photon polarization based on the vectors \eqref{eq:photon_polarization_x_z}. Like in Fig. \ref{fig:differential_cross_section_x_y} we have $\epsilon_f \in \{ \Le, \Ri \}$ and 
  $s_f \in \{ \searrow,\nwarrow \}$ for the summation and the plot in panel (a) is repeated on a logarithmic $y$-axis in panel (b) and is zoomed in panel (c) for resolving the properties of the cross section at $\vartheta=\pi$. Also the line styles are the same as in Fig. \ref{fig:differential_cross_section_x_y}. The dip at $\vartheta=\pi$ has a very similar shape as in the $x$-$y$ plane. However, in the $x$-$z$ plane one can see another dip at $\vartheta\approx0.4\pi$.
  \label{fig:differential_cross_section_x_z}
}%
\end{figure}

For a photon, which is scattered into the $x$-$z$ plane we set $\varphi=\pi/2$ and evaluate the Compton tensor \eqref{eq:compton_tensor} with the polarization vectors
\begin{subequations}%
\begin{align}%
 \epsilon^{\prime\Ve}(\vartheta) &= \vec e_y \\
 \epsilon^{\prime\Ho}(\vartheta) &= \sin \vartheta \,\vec e_x - \cos \vartheta \,\vec e_z\,,
\end{align}\label{eq:photon_polarization_x_z}%
\end{subequations}%
for which \eqref{eq:polarization_vector_orthogonality} is fulfilled as well. The polarizations in Eq. \eqref{eq:photon_polarization_x_z}, $k'$ and $p_f$ are also coincident with the definitions made in section \ref{sec:compton_scattering} at scattering angle $\theta=\pi$ and $\varphi=\pi/2$. Figure \ref{fig:differential_cross_section_x_z} shows the differential cross section in the interval $\vartheta \in [0,2\pi]$ in the $x$-$z$, plane similarly to the plot in the $x$-$y$ plane in Fig. \ref{fig:differential_cross_section_x_y}. The cross section in Fig. \ref{fig:differential_cross_section_x_z} also has a dip at $\vartheta=\pi$, which agrees with the findings in section \ref{sec:compton_scattering} and equation \eqref{eq:Vse-Lnw_scattering}, stating that an incoming state $\ket{\Ve,\searrow}$ is scattered in the final state $\ket{\Le,\nwarrow}$, if the photon is backstattered by 180 degrees. One can also see another dip in the cross section in the $x$-$z$ plane at $\vartheta\approx0.4\pi$.

\subsection{Conditional Stokes parameters}

We further investigate the electron-photon scattering with the Stokes parameters \cite{jackson_1962_classical_electrodynamics}
\begin{subequations}%
\begin{align}%
 \Pi_0 &= P_{\Ho,\nwarrow} + P_{\Ve,\nwarrow}\\
 \Pi_1 &= P_{\Ho,\nwarrow} - P_{\Ve,\nwarrow}\\
 \Pi_2 &= P_{\Di,\nwarrow} - P_{\An,\nwarrow}\\
 \Pi_3 &= P_{\Le,\nwarrow} - P_{\Ri,\nwarrow}\,,
\end{align}\label{eq:stokes_parameters}%
\end{subequations}%
with the absolute value squares of a normalized projection
\begin{equation}
 P_{\epsilon_f,s_f} = \left|\frac{\braket{\epsilon_f,s_f|\psi_f}}{\braket{\psi_f|\psi_f}}\right|^2\,.
\end{equation}

\begin{figure}[!h]%
  \includegraphics[width=0.48\textwidth]{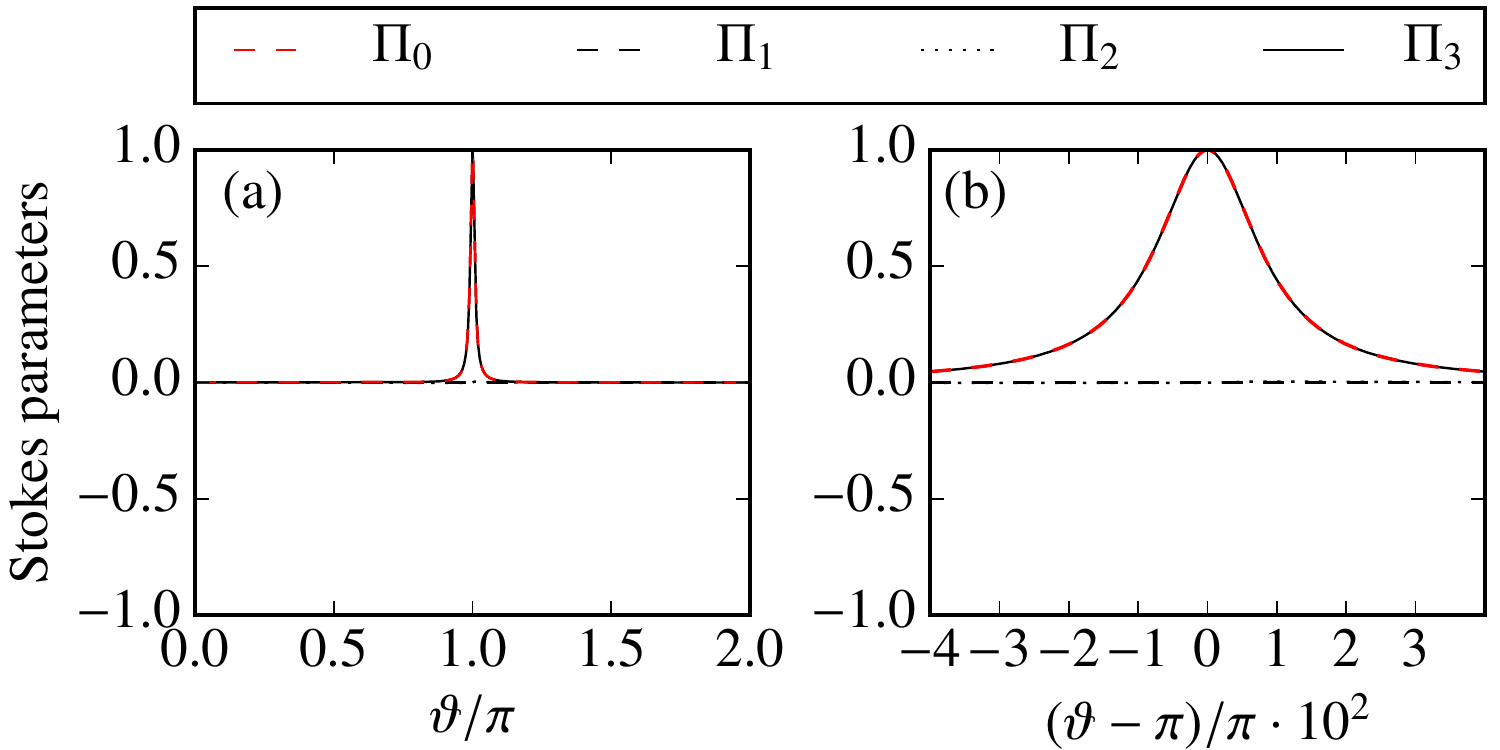}%
  \caption{(Color online) Conditional Stokes parameters \eqref{eq:stokes_parameters} in the $x$-$y$ plane for the final electron spin orientation $s^\nwarrow$ along the final photon momentum \eqref{eq:kp} with $\varphi=0$. Panel (a) is plotting on the interval $\vartheta \in [0,2\pi]$, where panel (b) zooms in the peaked region at $\vartheta=\pi$. One can see a sharp peak of the Stokes parameter $\Pi_3$ (solid black line) at $\vartheta=\pi$, coincident with the parameter $\Pi_0$ (red dashed line). The parameters $\Pi_1$ and $\Pi_2$ are zero, implying that the polarization peak is purely left-circularly polarized.
  \label{fig:stokes_parameters_x_y}}%
\end{figure}

\begin{figure}[!h]%
  \includegraphics[width=0.48\textwidth]{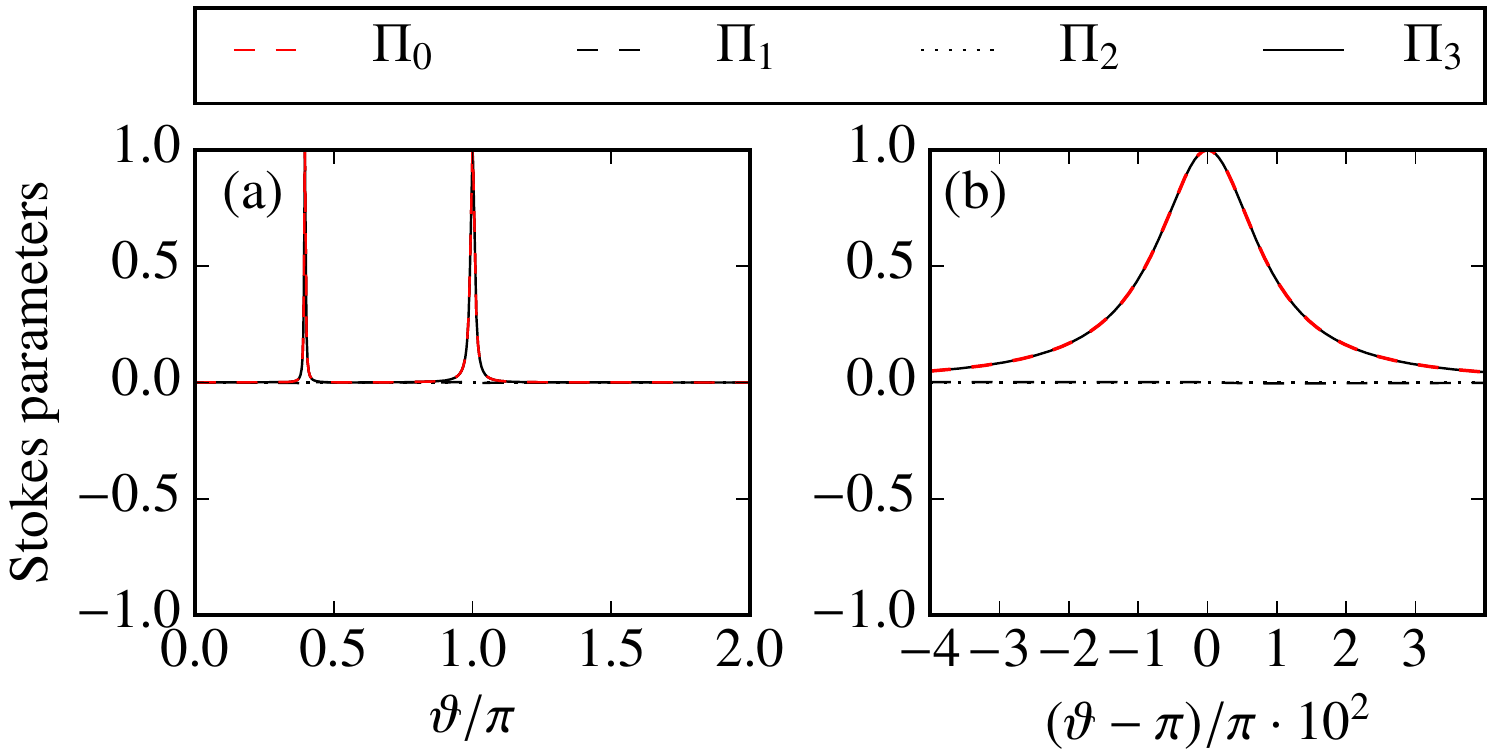}%
  \caption{(Color online) Conditional Stokes parameters \eqref{eq:stokes_parameters} in the $x$-$z$ plane for the final electron spin orientation $s^\nwarrow$ with $\varphi=\pi/2$. The line styles and $x$-axis are the same as in Fig. \ref{fig:stokes_parameters_x_y}. Like in Fig. \ref{fig:stokes_parameters_x_y} one concludes purely left-circularly polarized polarization. Beside the peak at $\vartheta=\pi$, there is another peak at location $\vartheta=0.4\pi$, corresponding to the second dip of the differential cross section in the $x$-$z$ plane, see Fig. \ref{fig:differential_cross_section_x_z}.
  \label{fig:stokes_parameters_x_z}
}%
\end{figure}

We mention that common labelings for the Stokes parameters are $(s_0,s_1,s_2,s_3)$ or $(I,Q,U,V)$, see \cite{jackson_1962_classical_electrodynamics}. Since we use $s$ already for denoting the electron spin and V for denoting vertical photon polarization, we introduce the new labels $(\Pi_0,\Pi_1,\Pi_2,\Pi_3)$ for the Stokes parameters in this article. In the framework of definition \eqref{eq:stokes_parameters}, the Stokes parameters are determined on the projection $\braket{\nwarrow|p_f}$ of the final scattering state $\ket{p_f}$ on the assumed electron spin state $\bra{\nwarrow}$\,. We plot the conditional Stokes parameters in the $x$-$y$ plane in Fig. \ref{fig:stokes_parameters_x_y} and in the $x$-$z$ plane in Fig. \ref{fig:stokes_parameters_x_z}. For the investigated case of $s^\nwarrow$ projected Stokes parameters one finds a sharp peak of the Stokes parameter $\Pi_3$ at $\vartheta=\pi$ in the $x$-$y$ plane and in the $x$-$z$ plane. This means that a generally unpolarized photon state is getting left-circularly polarized in the case of 180 degree back scattering of the photon, consistent with the findings in section \ref{sec:compton_scattering}.

\subsection{Summed Stokes parameters}

\begin{figure}[!h]%
  \includegraphics[width=0.48\textwidth]{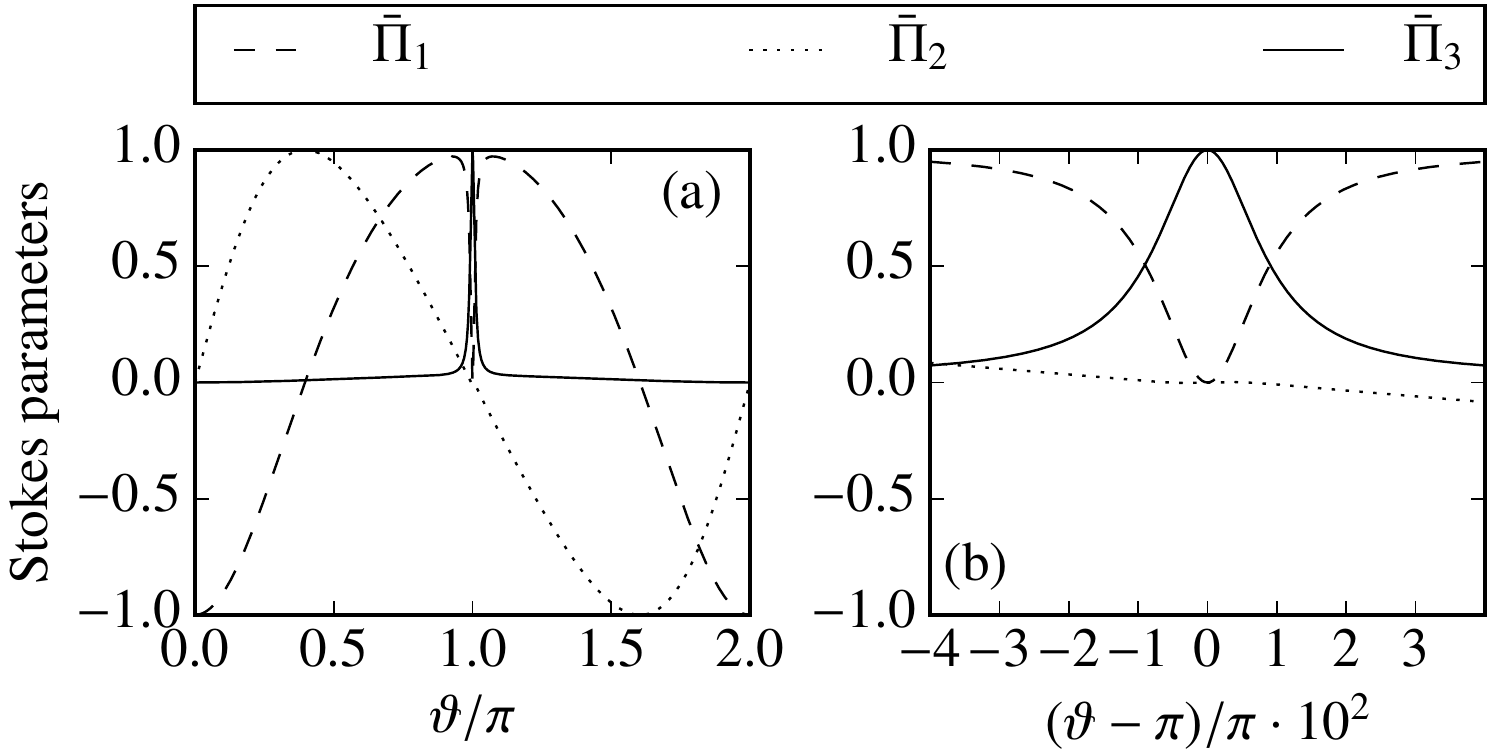}%
  \caption{Summed Stokes parameters \eqref{eq:summed_stokes_parameters} in the $x$-$y$ plane along the final photon momentum \eqref{eq:kp} with $\varphi=0$. Line style and $x$-axis are the same as for the conditional Stokes parameters in Fig. \ref{fig:stokes_parameters_x_y}. The parameter $\bar \Pi_0$ is not plotted, because of the general property $\bar \Pi_0=1$. One can see that the final photon is only circularly polarized at $\vartheta=\pi$ and linearly polarized elsewise. Here the photon polarization is vertically polarized at $\vartheta=0$ then changes into diagonal polarization at about $\vartheta\approx0.4\pi$, reaches horizontal polarization at around $\vartheta\approx\pi$ (except at the narrow peak at $\vartheta=0$) then changes into anti-diagonal polarization at about $\vartheta\approx1.6\pi$ and finally returns back to vertical polarization at $\vartheta=2 \pi$, as one sweeps around a circle in the $x$-$y$ plane.
  \label{fig:summed_stokes_parameters_x_y}
}%
\end{figure}

\begin{figure}[!h]%
  \includegraphics[width=0.48\textwidth]{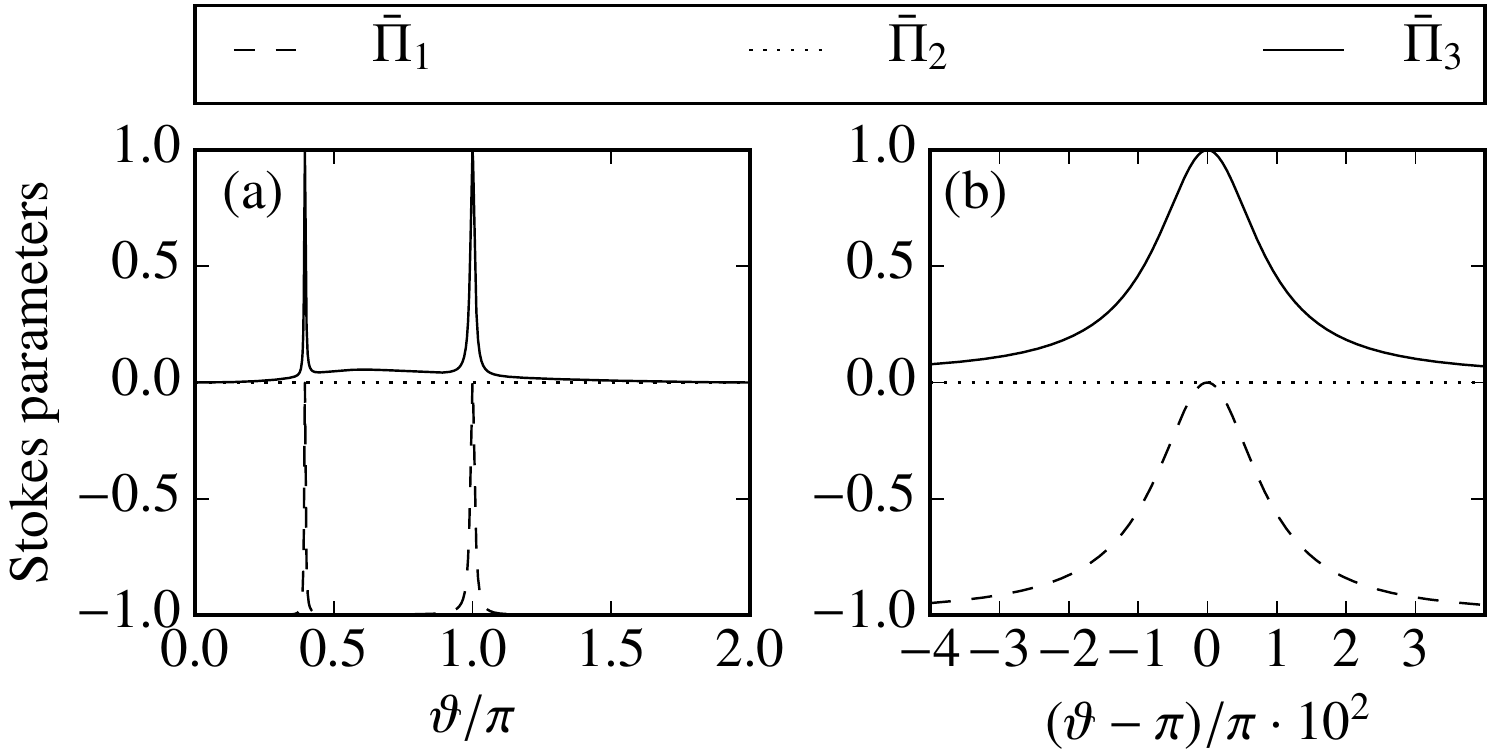}%
  \caption{Summed Stokes parameters \eqref{eq:summed_stokes_parameters} in the $x$-$z$ plane along the final photon momentum \eqref{eq:kp} with $\varphi=\pi/2$. Line style and $x$-axis are the same as in Fig. \ref{fig:summed_stokes_parameters_x_y} and $\bar \Pi_0$ is omitted as well. Again, we find two peaks with left-circular polarization at $\vartheta\approx0.4\pi$ and $\vartheta=\pi$ as for the conditional Stokes parameters in Fig. \ref{fig:stokes_parameters_x_z}. In contrast to diverse variations of the polarization in the $x$-$y$ plane in Fig. \ref{fig:summed_stokes_parameters_x_y}, the photon polarization just changes straight to horizontal polarization outside of the peak region.
  \label{fig:summed_stokes_parameters_x_z}
}%
\end{figure}

In order lift the constraint of the assumed electron spin projection state in Eq. \eqref{eq:stokes_parameters} we sum over the electron spin degree of freedom, resulting in the definition of the summed Stokes parameters
\begin{subequations}%
\begin{align}%
 \bar \Pi_0 &= P_{\Ho,\nwarrow} + P_{\Ho,\searrow} + P_{\Ve,\nwarrow} + P_{\Ve,\searrow} = 1\\
 \bar \Pi_1 &= P_{\Ho,\nwarrow} + P_{\Ho,\searrow} - P_{\Ve,\nwarrow} - P_{\Ve,\searrow}\\
 \bar \Pi_2 &= P_{\Di,\nwarrow} + P_{\Di,\searrow} - P_{\An,\nwarrow} - P_{\An,\searrow}\\
 \bar \Pi_3 &= P_{\Le,\nwarrow} + P_{\Le,\searrow} - P_{\Ri,\nwarrow} - P_{\Ri,\searrow}\,.
\end{align}\label{eq:summed_stokes_parameters}%
\end{subequations}%
We show the summed Stokes parameters in Fig. \ref{fig:summed_stokes_parameters_x_y} in the $x$-$y$ plane and Fig. \ref{fig:summed_stokes_parameters_x_z} in the $x$-$z$ plane. Linear polarization emerges for the final photon at locations apart of the peaks at $\vartheta\approx0.4\pi$ and $\vartheta=\pi$ in the $x$-$z$ plane. Since the linear polarization does not appear in the conditional Stokes parameters in Figs. \ref{fig:stokes_parameters_x_y} and \ref{fig:stokes_parameters_x_z}, the linear polarization emerges with a $\searrow$ electron polarization.

\subsection{Conditional and summed spin expectation value}

\begin{figure}[!h]%
  \includegraphics[width=0.48\textwidth]{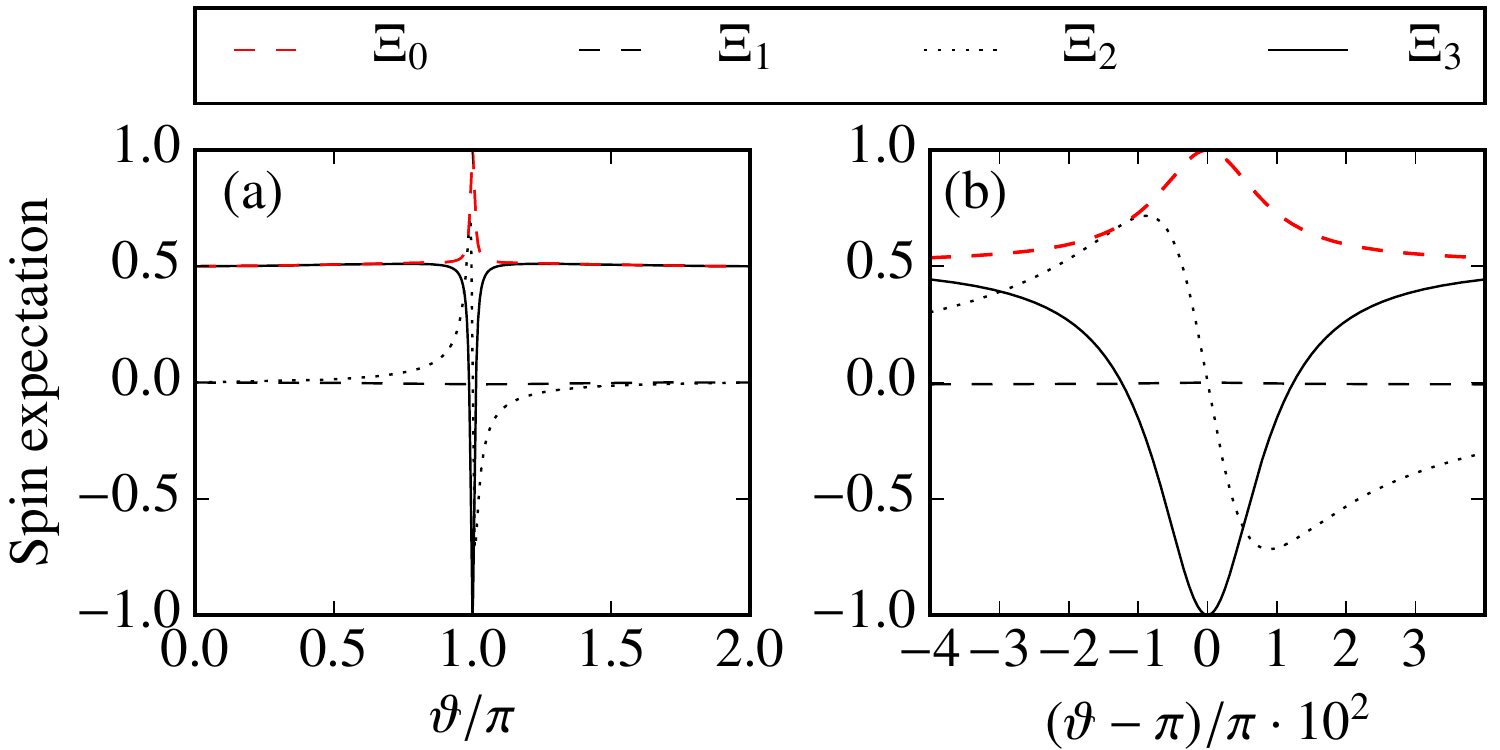}%
  \caption{(Color online) Conditional spin expectation \eqref{eq:spin_expectation} in the $x$-$y$ plane for a left circular final photon polarization plotted in the same way as the conditional Stokes parameters in Fig. \ref{fig:stokes_parameters_x_y}. Similar to the differential cross section and the Stokes parameters there is a peak of the parameter $\Xi_3$ (black solid line) at $\vartheta=\pi$, where it turns from 0.5 everywhere else into -1. This implies that the final electron spin is in a $\ket{\nwarrow}$ state at the peak.
  \label{fig:spin_expectation_x_y}
}%
\end{figure}

\begin{figure}[!h]%
  \includegraphics[width=0.48\textwidth]{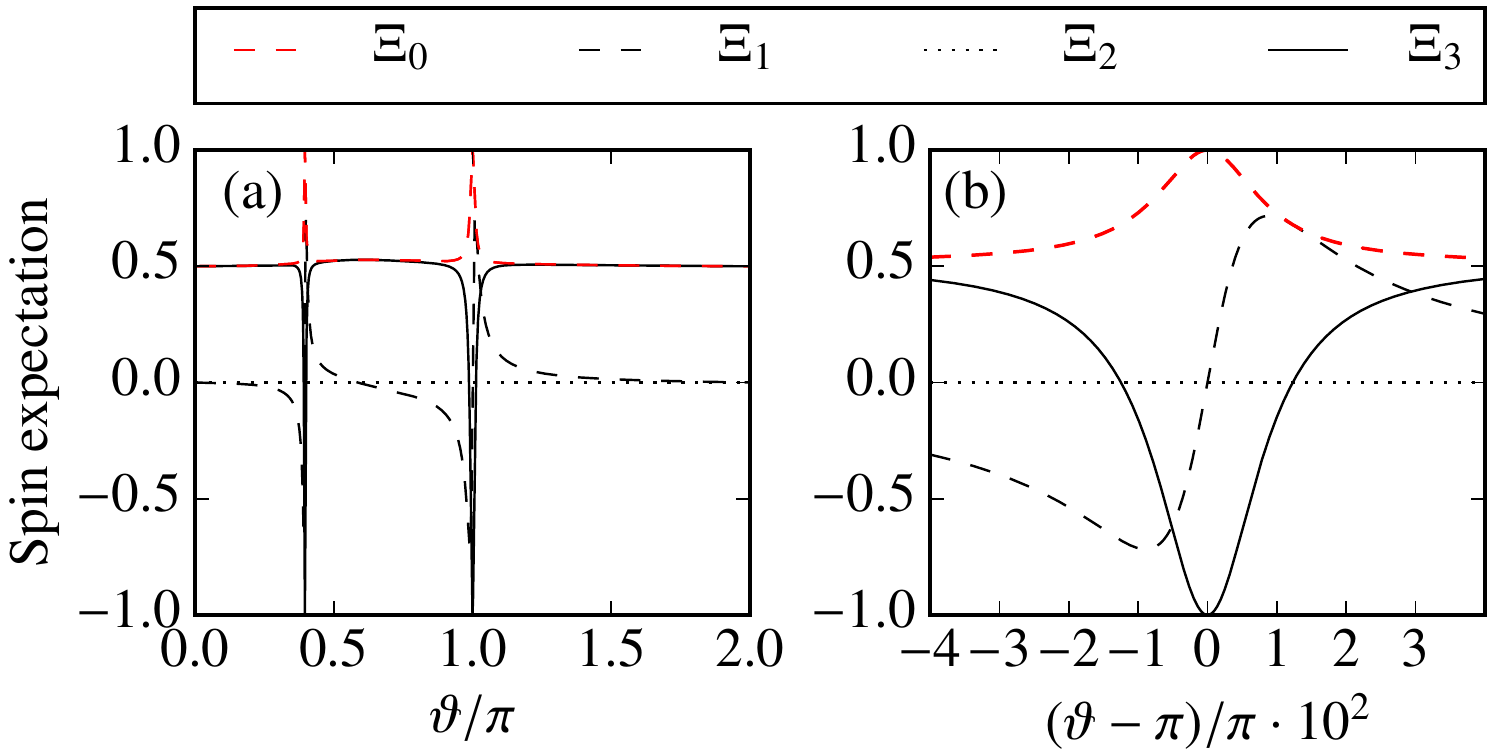}%
  \caption{(Color online) Conditional spin expectation \eqref{eq:spin_expectation} in the $x$-$z$ plane for a left circular final photon polarization plotted in the same way as the summed Stokes parameters in Fig. \ref{fig:stokes_parameters_x_z}. Similar to the 
  differential cross section and the Stokes parameters there is a second peak of $\Xi_3$ at $\vartheta\approx 0.4\pi$ in the $x$-$z$ plane.
  \label{fig:spin_expectation_x_z}
}%
\end{figure}

\begin{figure}[!h]%
  \includegraphics[width=0.48\textwidth]{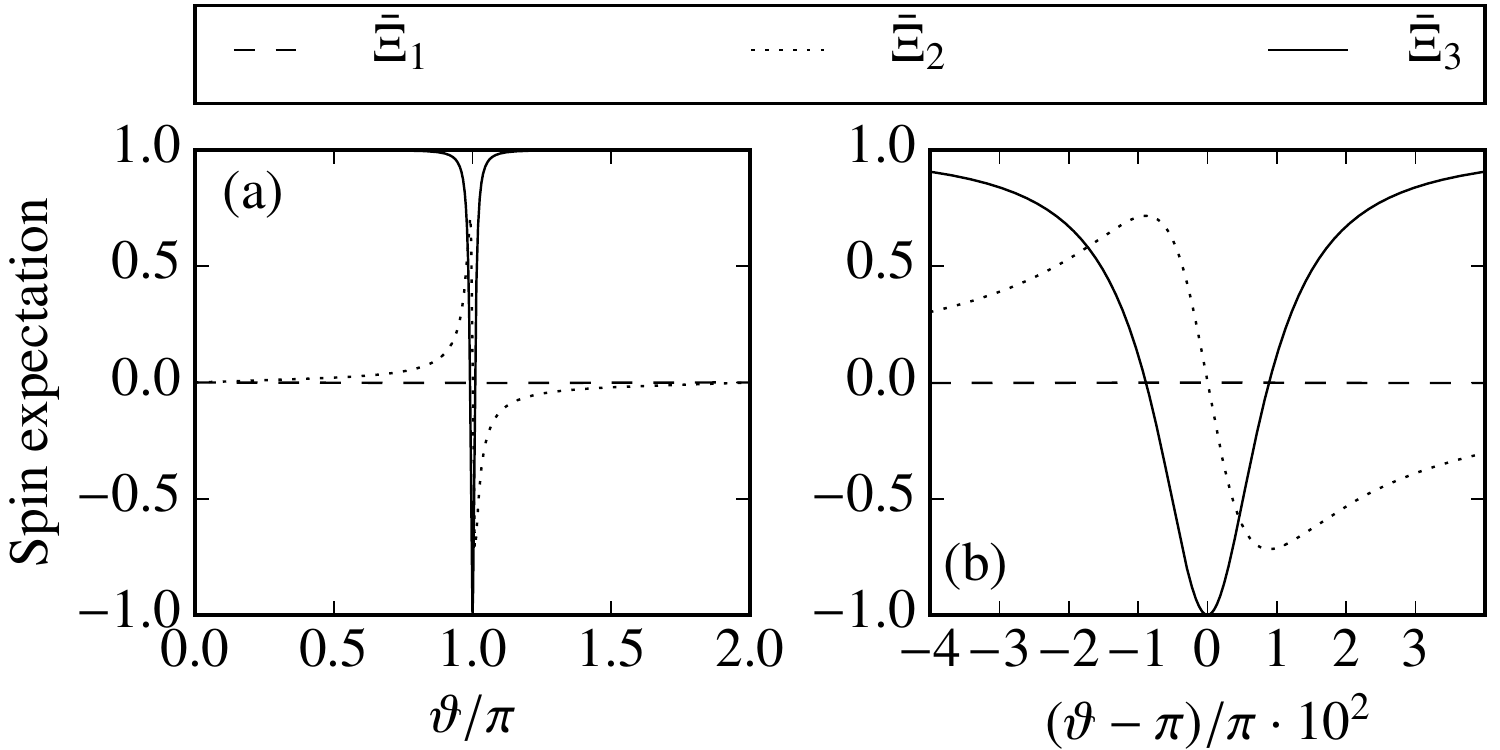}%
  \caption{Summed spin expectation \eqref{eq:summed_spin_expectation} in the $x$-$y$ plane, plotted in the same way as the summed Stokes parameters in Fig. \ref{fig:summed_stokes_parameters_x_y}. The angular spin dependence is qualitatively similar to the conditional spin expectation values in Fig. \ref{fig:spin_expectation_x_y}. However quantitatively, the parameter $\bar \Xi_3$ starts from the value 1 and changes over to -1 at the peak.\label{fig:summed_spin_expectation_x_y}
}%
\end{figure}

\begin{figure}[!h]%
  \includegraphics[width=0.48\textwidth]{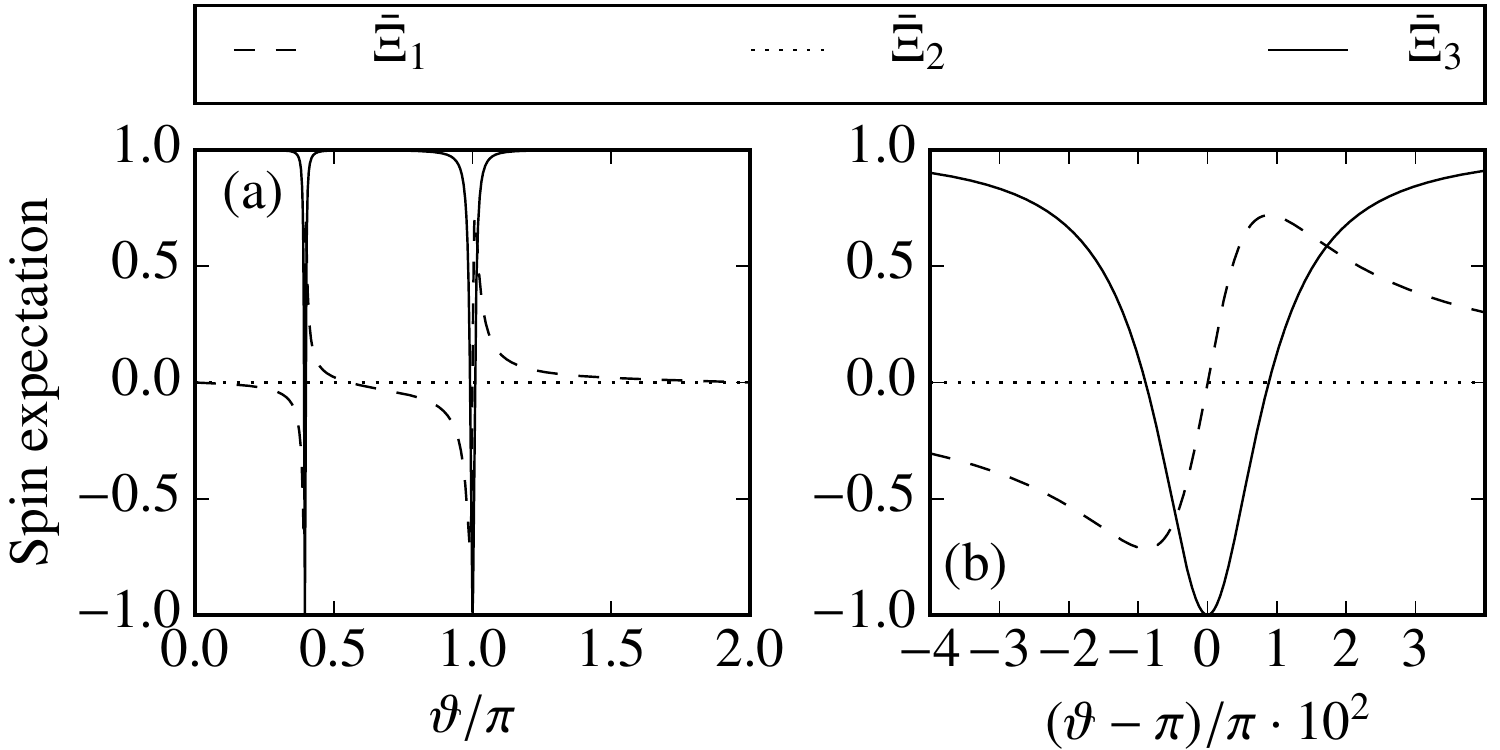}%
  \caption{Summed spin expectation \eqref{eq:summed_spin_expectation} in the $x$-$z$ plane, plotted in the same way as the conditional Stokes parameters in Fig. \ref{fig:summed_stokes_parameters_x_z}. Similar to the conditional spin expectation in Fig. \ref{fig:spin_expectation_x_z} one can see two peaks in the $x$-$z$ plane and analogously to Fig. \ref{fig:summed_spin_expectation_x_y}, the parameter $\bar \Xi_3$ starts from $1$ and goes down to $-1$ in the summed spin expectation.
  \label{fig:summed_spin_expectation_x_z}
}%
\end{figure}

Analogously to the conditional Stokes parameters in Eq. \eqref{eq:stokes_parameters} one can define symmetric parameters for the spin expectation value of the electrons
\begin{subequations}%
\begin{align}%
 \Xi_0 &= P_{\Le,\swarrow} + P_{\Le,\nearrow}\\
 \Xi_1 &= P_{\Le,\swarrow} - P_{\Le,\nearrow}\\
 \Xi_2 &= P_{\Le,\otimes} - P_{\Le,\odot}\\
 \Xi_3 &= P_{\Le,\searrow} - P_{\Le,\nwarrow}\,,
\end{align}\label{eq:spin_expectation}%
\end{subequations}%
with an assumed left-circularly polarized final photon state. And analogously to the summed Stokes parameters in Eq. \eqref{eq:summed_stokes_parameters} one can define a polarization summed electron spin expectation value
\begin{subequations}%
\begin{align}%
 \bar \Xi_0 &= P_{\Le,\swarrow} + P_{\Ri,\swarrow} + P_{\Le,\nearrow} + P_{\Ri,\nearrow} = 1\\
 \bar \Xi_1 &= P_{\Le,\swarrow} + P_{\Ri,\swarrow} - P_{\Le,\nearrow} - P_{\Ri,\nearrow}\\
 \bar \Xi_2 &= P_{\Le,\otimes} + P_{\Ri,\otimes} - P_{\Le,\odot} - P_{\Ri,\odot}\\
 \bar \Xi_3 &= P_{\Le,\searrow} + P_{\Ri,\searrow} - P_{\Le,\nwarrow} - P_{\Ri,\nwarrow}\,.
\end{align}\label{eq:summed_spin_expectation}%
\end{subequations}%
We plot the conditional spin expectation value \eqref{eq:spin_expectation} in Fig. \ref{fig:spin_expectation_x_y} in the $x$-$y$ plane and in Fig. \ref{fig:spin_expectation_x_z} in the $x$-$z$ plane. The polarization summed spin expectation value is plotted in Fig. \ref{fig:summed_spin_expectation_x_y} in the $x$-$y$ plane and in Fig. \ref{fig:summed_spin_expectation_x_z} in the $x$-$z$ plane. In contrast to the Stokes parameters, the plots of the conditional spin expectation values are qualitatively similar to the summed spin expectation values. However, the spin expectation values and the Stokes parameters have in common that first the parameter $\Xi_3$ has one peak in the $x$-$y$ plane and two peaks in the $x$-$z$ plane. Second similarity is that one of these peaks is located at $\vartheta=\pi$ in the $x$-$y$ plane as well as in the $x$-$z$ plane. And third common property is that the parameter $\Xi_3$ reaches the value -1 at the peak at $\vartheta=\pi$. Thus consistently with the Stokes parameters, the electron spin changes into a $\ket{\searrow}$ state which once more confirms the results in section \ref{sec:compton_scattering}.

\section{Experimental setup for an observation of the process\label{sec:experimental_setup}}

\begin{figure}[]%
  \includegraphics[width=0.48\textwidth]{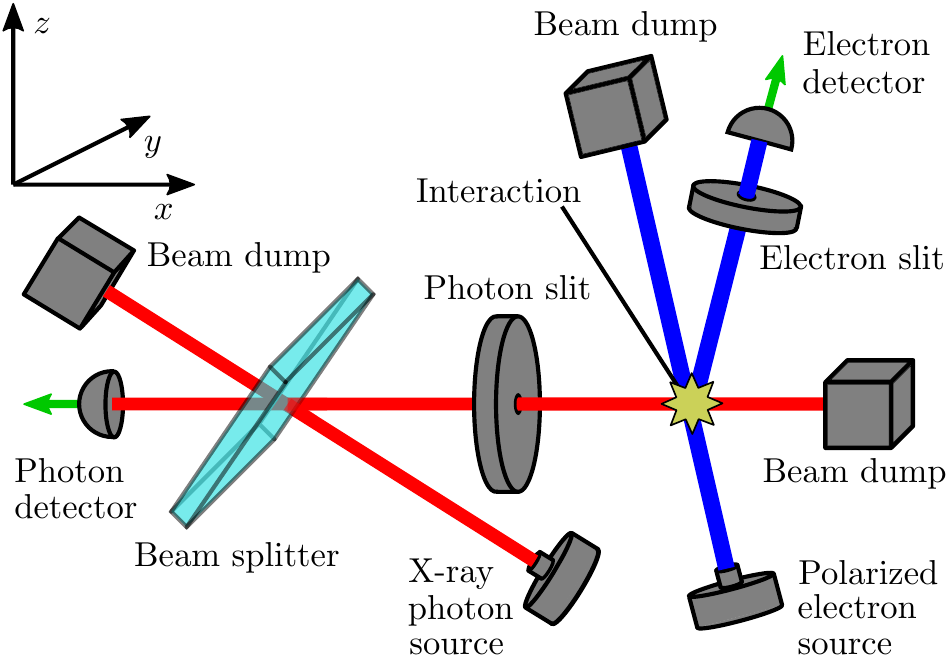}%
  \caption{\label{fig:experimental_setup}%
(Color online) Experimental setup for observing a spin non-preserving electron-photon interaction in Compton scattering. A 10.2\,keV photon is emitted from the X-ray source and is reflected \cite{Shvydko_2010_High-reflectivity_in_diamonds} by the beam splitter through the photon slit, such that it reaches the interaction region vertically polarized. In the interaction region, the photon scatters at a %$\sqrt{2}m$
212\,keV electron with polarization $s^\searrow$ from the polarized electron source. A coincident detection of a scattered electron and photon at the detectors corresponds to the process \eqref{eq:Vse-Lnw_scattering} with final electron spin $s^\nwarrow$ and a left-circularly polarized outgoing photon.
}%
\end{figure}

We suggest the observation of the discussed spin non-conserving dynamics in a scattering experiment which is sketched in Fig. \ref{fig:experimental_setup}.
In this scenario we assume the initial and final electron and photon momenta $p_i$, $p_f$, $k$ and $k'$ as described in section \ref{sec:compton_scattering} with the photon momentum $k_p=0.02 m$, corresponding to 10.2\,keV. The electron momentum component $p_2$ shall be zero and the electron momentum $p_3$ shall be such that the trace of $M_{33}$ is zero, resulting in $p_3\approx m$ 
\footnote{The Taylor expansion of $M_{33}$ indicates that $\textrm{tr}(M_{33})$ is not zero at the exact value $p_3=m$. We find by numeric evaluation of $M_{33}$ that its trace vanishes for $p_3\approx (1+8.2 \cdot 10^{-5}) m$ at $k_p=0.02 m$, which is the value of choice for our numeric evaluation of the Compton tensor \eqref{eq:compton_tensor}.}, implying the incoming electron's kinetic energy $\mathcal{E}_i-m\approx212\,\textrm{keV}$. A numeric evaluation \footnote{A python script which evaluates the Compton tensor Eq. \eqref{eq:compton_tensor} and obtains the numbers in Eq. \eqref{eq:xM} is enclosed in the supplementary information.} of the Compton tensor \eqref{eq:compton_tensor} for an incoming, vertically polarized photon and an electron with spin $s^\searrow$ polarization yields the transition amplitudes
\begin{subequations}%
\begin{align}%
 \left|\braket{\Le,\searrow|M|\Ve,\searrow}\right|^2&\approx 1.72\cdot10^{-13}\\
 \left|\braket{\Le,\nwarrow|M|\Ve,\searrow}\right|^2&\approx 4.00\cdot10^{-4}\label{eq:LuwM}\\
 \left|\braket{\Ri,\searrow|M|\Ve,\searrow}\right|^2&\approx 1.72\cdot10^{-13}\\
 \left|\braket{\Ri,\nwarrow|M|\Ve,\searrow}\right|^2&\approx  2.94\cdot10^{-14}\,,
\end{align}\label{eq:xM}%
\end{subequations}%
where the transition amplitude $\left|\braket{\Le,\nwarrow|M|\Ve,\searrow}\right|^2$ in Eq. \eqref{eq:LuwM} is consistent with the calculated process in Eq. \eqref{eq:Vse-Lnw_scattering}. The transition amplitudes into other polarizations than $\bra{\Le,\nwarrow}$ are negligibly small. This implies that each time when the electron detector and the photon detector of the setup in Fig. \ref{fig:experimental_setup} are receiving coincident signals, a vertically polarized photon and an electron with spin $s^\searrow$ are scattered into a left circularly polarized photon and an electron with spin $s^\nwarrow$, leaving a spin discrepancy of $1-1/\sqrt{2}$ in $x$ direction behind.

The differential cross section \eqref{eq:differential_cross_section} for a transition as in Eq. \eqref{eq:LuwM} has the value $12\,\mu$b. For $5\cdot 10^{16}$ photons per second from the photon source and $10^{17}$ electrons per second from the electron source one would expect only one electron and only one photon in the interaction region, if the photon beam had a spherical focus of $2\,\textrm{nm}$ diameter \cite{Schroer_Lengeler_2005_focussing_hard_X_rays} and the electron beam had an elliptical beam focus of $6\,\textrm{nm}\times700\,\textrm{nm}$ \cite{Yan_Oroku_2012_Nanometer_Beam_Size}. Having one electron in the assumed beam spot area of $4\,\textrm{nm}^2$ results in a collision probability of $6\cdot10^{-15}$ or 16 events per second for the considered photon flux. We further assume an uncertainty for the incoming electron and photon momenta below $1\,\textrm{keV}$. With this requirement, one can conclude from the Taylor expansion in Eq. \eqref{eq:taylor_expansion_compton_tensor} with respect to $k_p$, $p_2$ and $p_3$ that the process \eqref{eq:LuwM} will dominate over the other processes in Eq. \eqref{eq:xM} by one order of magnitude \footnote{In Eq. \eqref{eq:xM} we show the best case scenario, in which all processes except the process \eqref{eq:LuwM} are negligible, where a finite value on the order of $10^{-13}$ is attributed to higher orders in the powers $\alpha$ and $\gamma$ in the Taylor expansion \eqref{eq:taylor_expansion_compton_tensor} of the Compton tensor \eqref{eq:compton_tensor}. In the considered worst case, ie. a momentum deviation of $1\,\textrm{keV}/c$ out of $10\,\textrm{keV}/c$, one concludes that other contributions are about one order of magnitude lower than the leading contribution \eqref{eq:LuwM} from the form of the Taylor expansion \eqref{eq:taylor_expansion_compton_tensor}.}.

We mention that the transition amplitudes in \eqref{eq:xM} are only weakly depending on the initial electron momentum $p_1$, as shown in appendix \ref{sec:p1_numeric}. We discuss the dependence of all matrix elements of the Compton tensor \eqref{eq:compton_tensor} on $p_1$ by evaluating it in a new frame of reference in which $p_1 = - k_p$ holds in appendix \ref{sec:momentum_deviation}. This is achieved by choosing an appropriate Lorentz transformation in $x$ direction. The properties in the new frame of reference can be related to the lab frame, which is considered here, from which we conclude that for small momentum variations $|\delta p_1|\ll 1\,\textrm{keV}/c$ the matrix element \eqref{eq:LuwM} is dominating over the other transition amplitudes in Eq. \eqref{eq:xM}.

We also want to restrict the experiment to the peak of the Stokes parameters and the spin expectation value at $\vartheta=\pi$, as given in section \ref{sec:angular_analysis}, where the photon is back scattered by 180 degrees. According to the considerations in section \ref{sec:electron_and_photon_spin}, an intrinsic angular momentum of $(1 - 1/\sqrt{2})\hbar \approx 0.3\hbar$ is unexplained, when such an event occurs. In order to support this claim in the case of an event detection, we only want to allow final photon directions, for which the Stokes parameter $\Pi_3$ of the final photon is larger than 0.85 and the spin expectation value $\Xi_3$ of the final electron is below -0.7\,. In this case, the left-circular polarized photon generates $1\hbar$ intrinsic angular momentum in $x$ direction with more than $85\%$ likelihood and the electron spin flip generates $1\hbar$ intrinsic angular momentum in $(1,0,-1)^\T/\sqrt{2}$ direction with more than $85\%$ likelihood, such that the discrepancy of $0.3\hbar$ intrinsic angular momentum is surpassed along the $x$ axis on average. In Figs. \ref{fig:stokes_parameters_x_y}, \ref{fig:stokes_parameters_x_z}, \ref{fig:spin_expectation_x_y} and \ref{fig:spin_expectation_x_z} we see that $\Pi_3>0.85$ and $\Xi_3<-0.7$ is fulfilled for $- 5\cdot 10^{-3} \pi \le \vartheta - \pi \le 5\cdot 10^{-3} \pi$ for $\varphi=0$, along the $y$ axis, as well as for $\varphi=\pi/2$, along the $z$ axis. Thus, the considered spin non-conserving event in Compton scattering can be observed by choosing the spherical photon slit in the experiment in Fig. \ref{fig:experimental_setup}, such that only outgoing photons with a divergence of $5 \pi\,$mrad from the $x$ axis will reach the photon detector. With the resulting solid angle of $d\Omega=775\,\textrm{mrad}^2$ one expects about 1 photon every 160 seconds from a process of the form \eqref{eq:LuwM}, at a beam splitter's transmittivity of 50\%.

For the generation of entanglement as discussed in section \ref{sec:entanglement_generation} a beam splitter is not necessarily required in the experimental setup. The beam splitter is introduced for enabling the possibility of 180 degree back scattering of the interacting photon such that the longitudinal photon spin can be determined according to Eq. \eqref{eq:photon_spin}. The generation of polarization entanglement however, does not require co-aligned propagation directions for the incoming and outgoing photon, such that entanglement generation could also be implemented in a different frame of reference.

\section{Discussions and Outlook}

The investigated process in Eq. \eqref{eq:LuwM} leaves open questions. For example only those events, in which the photon is back-scattered are detected and accounted for. But still scattering events with different outgoing electron and photon momenta are taking place as well, see section \ref{sec:angular_analysis}. This demands for a study of the spin density integrated over the full angular dependence of the outgoing electron and photon momenta.

Also, we only discuss the intrinsic angular momentum (spin) of the particles and completely ignore the orbital angular momentum in our discussion. The reason is, that the orbital angular momentum of a plane wave integrated through a sphere vanishes. In reality, however, the electron and photon wave functions have finite extensions. Therefore, the angular momentum splitting into intrinsic and external angular momentum of the photon should be investigated with respect to finite and infinite extensions. In this context the study of orbital momentum eigensolutions for example of twisted light \cite{Molina-Terriza_2007_twisted_photons} or twisted electron states \cite{Bialynicki-Birula_2017_Relativistic_Electron_Angular_Momentum} and their interaction could also be interesting. Steps in this direction were undertaken recently based on plane wave electrons, twisted light and a non-relativistic interaction \cite{Stock_2015_Compton_scattering_of_twisted_light}.

What is also of interest is the interaction of a similar process, in which the electron is interacting with the strong field versions of the incoming and outgoing photons, similar to \cite{ahrens_bauke_2013_relativistic_KDE}. In this case the electron wavefunction is expected to perform deterministic dynamics, due to absorption and simulated emission in the strong fields (provided the fields are coherent as for example in laser beams). However, then it is more of interest, how the electron is interacting with the many particle quantum state of the light field. We expect that the polarization properties of each of the photon number states in the laser beam's coherent light field will be influenced by the electron in a non-trivial way.

\begin{acknowledgments}
S.A. thanks for discussions with Tilen Cadez, Carsten M\"uller, Karen Z. Hatsagortsyan, Enderalp Yakaboylu, Heiko Bauke and Sebastian Meuren. This work has been supported by the National Basic Research Program of China (Grant No. 2016YFA0301201 and No. 2014CB921403), by the NSFC (Grant No. 11650110442 and No. 11421063 and No. 11534002) and the NSAF (Grant No. U1530401).
\end{acknowledgments}

\appendix

\section{Equivalent spinor expressions\label{sec:spinor_expressions}}

The spinors $s^\searrow$ and $s^\nwarrow$ given in Eq. \eqref{eq:tilted_spinor_definition} can also be denoted by the equivalent expressions
\begin{subequations}
\begin{align}
 s^\searrow &= N_-
\begin{pmatrix}
  1 - \sqrt{2} \\ -1
\end{pmatrix} \\
 s^\nwarrow &= N_+
\begin{pmatrix}
 1 + \sqrt{2} \\ -1
\end{pmatrix}
\end{align}
\end{subequations}
with the normalization factors
\begin{subequations}
\begin{align}
N_+ &= \sqrt{2(2+\sqrt{2})}^{-1}\\
N_- &= \sqrt{2(2-\sqrt{2})}^{-1}\,.
\end{align}
\end{subequations}
We use these expressions in for the numeric evaluation of the Compton tensor \eqref{eq:compton_tensor}.

\section{Four momentum conservation and differential cross section\label{sec:differential_cross_section}}

The initial electron momentum  \eqref{eq:initial_electron_momentum} with $p_2=0$, the initial photon momentum \eqref{eq:initial_photon_momentum}, the final photon momentum \eqref{eq:kp} and the momentum conservation \eqref{eq:four_momentum_conservation} imply the final electron momentum \eqref{eq:final_electron_momentum}. Equation \eqref{eq:four_momentum_conservation} also implies energy conservation, which we write as
\begin{equation}
 \mathcal{E}_f = \mathcal{E}_i + k_p - \omega'\,.\label{eq:energy_conservation}
\end{equation}
The square of the left-hand side of Eq. \eqref{eq:energy_conservation} evaluates to
\begin{equation}
 \mathcal{E}_f^2 = m^2 + \omega^{\prime 2} + p_3^2 - 2 \omega' p_3 \sin \vartheta \sin \varphi\,.
\end{equation}
The square of the right-hand side results in
\begin{equation}
 (\mathcal{E}_i + k_p - \omega')^2 = m^2 + 2 k_p^2 + p_3^2 +  \omega'^2 -2 k_p \omega' + 2 \mathcal{E}_i k_p - 2 \mathcal{E}_i \omega'\,.
\end{equation}
Plugging these two terms into the square of equation \eqref{eq:energy_conservation} and solving for $\omega'$ results in the final photon momentum
\begin{equation}
 \omega' = k_p \frac{\mathcal{E}_i + k_p}{\mathcal{E}_i + k_p - p_3 \sin \vartheta \sin \varphi} \,. \label{eq:final_electron_momentum_appendix}
\end{equation}
We refer again to reference \cite{greiner_reinhardt_1992_quantum_electrodynamics} chapter 3.7, in which the cross section is given by
\begin{multline}
 d \sigma = \frac{q^4}{(2 \pi)^2} \frac{1}{m^2} \frac{m (4 \pi)^2}{\mathcal{E}_i |\Delta \vec v| 2 \omega} \int \delta^4(p_f + k' - p_i - k)\\
 \cdot|\epsilon^{\prime \mu} \,{M_{\mu\nu}}\, \epsilon^\nu|^2 
 \frac{m \,d^3 p_f}{\mathcal{E}_f} \frac{d^3 k'}{2 \omega'}\,.\label{eq:cross_section}
\end{multline}
The phase space volume integral over the four-dimensional delta function can be written as
\begin{multline}%
 \int \delta^4(p_f + k' - p_i - k) \frac{m \, d^3 p_f}{\mathcal{E}_f} \frac{d^3 k'}{2 \omega'} \\= m \int_0^{m + k_p} \delta[(p_i + k - k')^2 - m^2] \,\omega' d\omega' d\Omega \,,\label{eq:phase_space_integral}
\end{multline}%
where the argument of the delta function simplifies to
\begin{multline}
  (p_i + k - k')^2 - m^2 = 2 k_p^2 -2 k_p \omega' + 2 \mathcal{E}_i k_p - 2 \mathcal{E}_i \omega' \\
  + 2 \omega' p_3 \sin \vartheta \sin \varphi
\end{multline}
for the initial and final photon and electron momenta considered here. The derivative of this equation with respect to $\omega'$ yields
\begin{subequations}
\begin{align}
 \frac{\partial}{\partial \omega'}\left[(p_i + k - k')^2 - m^2 \right] &= -2 k_p - 2 \mathcal{E}_i + 2 p_3 \sin \vartheta \sin \varphi \nonumber \\
 &= - 2 k_p \frac{\mathcal{E}_i + k_p}{\omega'}\,,
\end{align}
\end{subequations}
with relation \eqref{eq:final_electron_momentum_appendix} being substituted. The phase space integral \eqref{eq:phase_space_integral} evaluates into
\begin{equation}%
 \int \delta^4(p_f + k' - p_i - k) \frac{m \, d^3 p_f}{\mathcal{E}_f} \frac{d^3 k'}{2 \omega'} =  \frac{\omega^{\prime2}}{2 k_p}\frac{m}{\mathcal{E}_i + k_p} d \Omega\,,\label{eq:phase_space_integral_evaluated}
\end{equation}%
where the integral identity for delta functions
\begin{equation}
 \int dx \, \delta[f(x)] \, g(x) = \sum_{\{x \in \mathbb{R},f(x)=0\}} g(x)\left| \frac{\partial f(x)}{\partial x} \right|^{-1}
\end{equation}
has been used. By inserting Eq. \eqref{eq:phase_space_integral_evaluated} in Eq. \eqref{eq:cross_section} and dividing by $d \Omega$ we arrive at Eq. \eqref{eq:differential_cross_section}. The absolute value of the relative particle velocity $\Delta \vec v = \vec v_\gamma - \vec v_e$ in Eq. \eqref{eq:cross_section} is substituted with
\begin{equation}
 |\Delta \vec v| = \mathcal{E}_i^{-1} \sqrt{\mathcal{E}_i^2 - 2 \mathcal{E}_i k_p + k_p^2 + p_3^2} \,,
\end{equation}
where $\vec v_\gamma = \vec e_x$ is the initial photon velocity and $\vec v_e = \vec p_i/\mathcal{E}_i$ is the initial electron velocity.

\section{Longitudinal momentum deviation}

\subsection{Analytic considerations\label{sec:momentum_deviation}}

\noindent In the main text a particle with initial momentum
\begin{equation}
 p_i=
 \begin{pmatrix}
  E(\vec p) \\ -k_p \\ p_2 \\ p_3
 \end{pmatrix}\label{eq:initial_electron_four_momentum}
\end{equation}
is considered to scatter a photon with initial momentum
\begin{equation}
 k=
 \begin{pmatrix}
  k_p \\ k_p \\ 0 \\ 0
 \end{pmatrix}\,,\label{eq:initial_photon_four_momentum}
\end{equation}
such that the momenta in $x$ direction are reversed after interaction. Assume the $x$ component of the incoming electron is changed by the small value $\delta p$ and changes into $-k_p + \delta p$ instead of $-k_p$. Then it is possible to change the frame of reference along the $x$ direction, such that the transformed $x$ component of the incoming electron's four momentum $\tilde p_x$ has the same value but opposite sign as the transformed $x$ component $\tilde k_x$ of the transformed four momentum of the incoming photon. For obtaining an explicit expression for such a transformation we express the incoming four momentum of the electron in terms of the rapidity (at least partially)
\begin{equation}
 p_i=
 \begin{pmatrix}
  \tilde E \cosh(\eta) \\ -\tilde E \sinh(\eta) \\ p_2 \\ p_3
 \end{pmatrix}\,.\label{eq:rapidity_electron_four_momentum}
%  \,,\qquad
%  k=
%  \begin{pmatrix}
%   e^{\eta'} \\ e^{\eta'} \\ 0 \\ 0
%  \end{pmatrix}
\end{equation}
The parameters $\tilde E$ and $\eta$ are related to the parameters $E(\vec p)$, $k_p$, $p_2$ and $p_3$ in Eq. \eqref{eq:initial_electron_four_momentum} and Eq. \eqref{eq:initial_photon_four_momentum}. The inner product $p_\mu p^\mu = m$ in Minkowski space implies that $\tilde E$ fulfills the relation
\begin{equation}
 \tilde E^2 = m^2 + p_2^2 + p_3^2
\end{equation}
and $\eta$ is related by
\begin{equation}
 \sinh(\eta) = \frac{k_p}{\tilde E}\,.\label{eq:eta_relation}
\end{equation}
The change of the electron momentum by $\delta p$ can be expressed by the Lorentz transformation
\begin{equation}
\Lambda(\Delta) =
\begin{pmatrix}
 \cosh(\Delta) & -\sinh(\Delta) & 0 & 0 \\
 -\sinh(\Delta) & \cosh(\Delta) & 0 & 0 \\
 0 & 0 & 1 & 0 \\
 0 & 0 & 0 & 1
\end{pmatrix}\label{eq:lorentz-transformation}
\end{equation}
and changes the electron four-vector into
\begin{equation}
 \Lambda(\Delta) p_i =
 \begin{pmatrix}
  \tilde E \cosh(\eta + \Delta) \\ -\tilde E \sinh(\eta + \Delta) \\ p_2 \\ p_3
 \end{pmatrix}\,.
\end{equation}
Similarly to Eq. \eqref{eq:eta_relation} one can write
\begin{equation}
 \sinh(\eta + \Delta) = \frac{k_p - \delta p}{\tilde E}\label{eq:eta_delta_relation}
\end{equation}
which establishes the relation between $\delta p$ and $\Delta$ in Eq. \eqref{eq:lorentz-transformation}.

According to the statement above we want to perform a Lorentz transformation of the system, such that the $x$ components of the electron and photon momenta are equal but have opposite sign, corresponding to the equations \eqref{eq:initial_electron_four_momentum} and \eqref{eq:initial_photon_four_momentum} before the momentum disturbance $\delta p$. Using the transformation \eqref{eq:lorentz-transformation} with the new transformation parameter $\theta$ changes the four vectors of the electron and the photon into
\begin{subequations}%
\begin{align}%
 \Lambda(\theta) \Lambda(\Delta) p_i &=
 \begin{pmatrix}
  \tilde E \cosh(\eta + \Delta + \theta) \\ -\tilde E \sinh(\eta + \Delta + \theta) \\ p_2 \\ p_3
 \end{pmatrix}\\
 \Lambda(\theta) k &=
 \begin{pmatrix}
  k_p e^{- \theta} \\ k_p e^{- \theta} \\ 0 \\ 0
 \end{pmatrix}\,.
\end{align}%
\end{subequations}%
As described above, the parameter $\theta$ of the Lorentz transformation has to be such that the $x$ component of the electron and photon momentum are equal but opposite, which implies that
\begin{equation}
 -\tilde E \sinh(\eta + \Delta + \theta) = k_p e^{- \theta}
\end{equation}
shall hold. Due to the equality before the change by the small momentum $\delta p$ in Eq. \eqref{eq:initial_electron_four_momentum} and Eq. \eqref{eq:initial_photon_four_momentum} and the equivalent expression Eq. \eqref{eq:rapidity_electron_four_momentum} for the electron momentum one can further substitute
\begin{equation}
 -\tilde E \sinh(\eta + \Delta + \theta) = -\tilde E \sinh(\eta) e^{- \theta}\,.
\end{equation}
The sine hyperbolic functions can be expanded in terms of exponential functions
\begin{equation}
 \frac{1}{2} e^{\eta + \Delta + \theta} - \frac{1}{2} e^{-\eta - \Delta - \theta} = \frac{1}{2} e^{\eta - \theta} - \frac{1}{2} e^{-\eta - \theta}
\end{equation}
and solved for $\theta$, resulting in
\begin{equation}
 \theta = \frac{1}{2} \ln\left( e^{-\Delta} - e^{-2 \eta -\Delta} + e^{-2 \eta -2 \Delta} \right)\,.
\end{equation}
A Taylor expansion with respect to $\Delta$ results in
\begin{equation}
 \theta = - \cosh(\eta) e^{-\eta} \Delta + \mathcal{O}(\Delta^2)\label{eq:taylor_delta}
\end{equation}
Also, plugging Eq. \eqref{eq:eta_relation} into Eq. \eqref{eq:eta_delta_relation} results in
\begin{equation}
 \sinh(\eta + \Delta) = \sinh(\eta) - \frac{\delta p}{\tilde E}\,.
\end{equation}
Solving for $\Delta$ results in
\begin{equation}
 \Delta = \arcsin\left( \sinh(\eta) - \frac{\delta p}{\tilde E} \right) - \eta\,,
\end{equation}
and a Taylor expansion with respect to $\eta$ yields
\begin{equation}
 \Delta = - \frac{\delta p}{\tilde E \cosh(\eta)} + \mathcal{O}(\eta^2)
\end{equation}
Plugging this into Eq. \eqref{eq:taylor_delta} results in
\begin{equation}
 \theta = \frac{e^{-\eta}}{\tilde E}\delta p + \mathcal{O}(\eta^2)\,.\label{eq:taylor_theta}
\end{equation}
We note that the Lorentz transformation \eqref{eq:lorentz-transformation} will only transform the longitudinal and time-like components of tensors, while transverse components are not affected. Thus the transverse components of the Compton tensor in Eq. \eqref{eq:compton_tensor} of the main text are not changed by the transformation \eqref{eq:lorentz-transformation}, except an implicit change of the photon momentum $k_p$ in the new frame of reference. However, since the right-hand side of \eqref{eq:taylor_theta} is assumed to be much smaller than $k_p/m$, the shifted photon momentum in the new frame of reference due to a small change $\delta p$ of the electron momentum in $x$ direction scales smaller than $k_p^2/m^2$ in Eq. \eqref{eq:photon_momenta} of the main text and is therefore negligible.

The reader may get the impression that a negligible change of the Compton tensor by momentum $\delta p$ only applies in the new frame of reference, but not in the lab frame, in which the actual interaction takes place as illustrated in Fig. \ref{fig:electron-photon_interaction}. But since the transverse components of Compton tensor are invariant under the Lorentz transformation \eqref{eq:lorentz-transformation}, the conclusion in the new frame of reference applies for the lab frame as well.

\subsection{Numeric investigation\label{sec:p1_numeric}}

\begin{figure}[!h]%
  \includegraphics[width=0.48\textwidth]{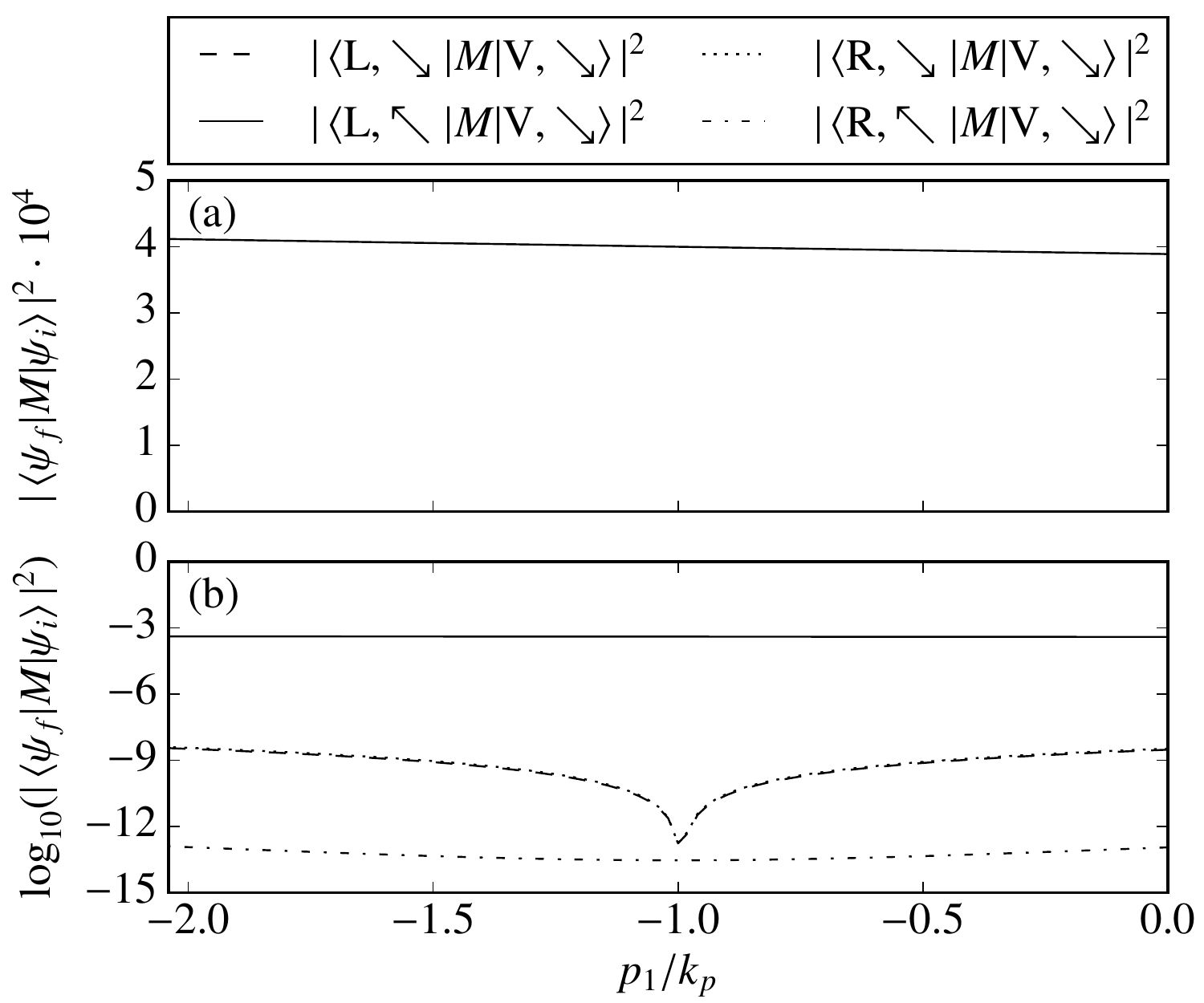}%
  \caption{Absolute value squares of the matrix elements of the Compton tensor \eqref{eq:compton_tensor}. In panel (a) one can see, that only $\left|\braket{\Le,\nwarrow|M|\Ve,\searrow}\right|^2$ has a value of about $4.00\cdot10^{-4}$, in accordance with Eq. \eqref{eq:LuwM}. The other matrix elements are smaller by several orders of magnitudes for a vast range of the momentum component $p_{1}$, as can be seen in the logarithmic plot in panel (b).
  \label{fig:projection_p1}
}%
\end{figure}

We want to explicitly check the transition amplitudes \eqref{eq:xM} for a variation of component $p_{1}$ of the initial electron momentum $\vec p_i$. Therefore we plot the transition amplitudes in Fig. \ref{fig:projection_p1}. The matrix element $\left|\braket{\Le,\nwarrow|M|\Ve,\searrow}\right|^2$ is dominating over the other final scattering configurations by several orders of magnitude, such that we conclude a negligible dependence of the momentum $p_{1}$ on the amplitudes (36).

\bibliography{bibliography}

\end{document}